\documentclass[a4paper,11pt]{article}
\pdfoutput=1 % if your are submitting a pdflatex (i.e. if you have
\usepackage{jinstpub-custom}

\usepackage{lineno}
\usepackage{siunitx}
\usepackage{subcaption}
\usepackage[noabbrev]{cleveref}
\usepackage{xspace}

\title{\boldmath Measurements of time resolution of the RD50-MPW2 DMAPS prototype using TCT and $\Sr$}

\author[a,1]{J.~Debevc,\note{Corresponding author.}}
\author[b,2]{M.~Franks,\note{Now at ETH Zürich.}}
\author[a]{B.~Hiti,}
\author[c]{U.~Kraemer,}
\author[a]{G.~Kramberger,}
\author[a]{I.~Mandić,}
\author[d]{R.~Marco-Hernández,}
\author[c]{D.J.L.~Nobels,}
\author[b]{S.~Powell,}
\author[c]{J.~Sonneveld,}
\author[e]{H.~Steininger,}
\author[c,3]{C.~Tsolanta,\note{Now at University of Oslo.}}
\author[b]{E.~Vilella}
\author[b]{and C.~Zhang}

\affiliation[a]{Jožef Stefan Institute,\\Jamova cesta 39, Ljubljana, Slovenia}
\affiliation[b]{University of Liverpool, Department of Physics,\\Oxford Street, Liverpool, United Kingdom}
\affiliation[c]{Nikhef,\\Science park 105, Amsterdam, Netherlands}
\affiliation[d]{IFIC (CSIC-UV),\\Valencia, Spain}
\affiliation[e]{HEPHY,\\Nikolsdorfer Gasse 18, Vienna, Austria}

\emailAdd{jernej.debevc@ijs.si}

\abstract{Results in this paper present an in-depth study of time resolution for active
pixels of the RD50-MPW2 prototype CMOS particle detector. Measurement techniques
employed include Backside- and Edge-TCT configurations, in addition to electrons
from a $\Sr$ source. A sample irradiated to \fluence was used to study the
effect of radiation damage. Timing performance was evaluated for the entire
pixel matrix and with positional sensitivity within individual pixels as a
function of the deposited charge. Time resolution obtained with TCT is seen to
be uniform throughout the pixel's central region with approx.\ \SI{220}{\ps} at
\SI{12}{\kilo\electron} of deposited charge, degrading at the edges and lower
values of deposited charge. $\Sr$ measurements show a slightly worse time
resolution as a result of delayed events coming from the peripheral areas of the
pixel.
}

\keywords{
	Solid state detectors, Radiation-hard detectors, Particle tracking detectors (Solid-state detectors)
}

\arxivnumber{2312.01793}

\newcommand{\Sr}{^{90}\mathrm{Sr}} % Strontium-90
\newcommand{\fluence}{\SI{5e14}{\neq\per\cm\squared}\xspace} % Fluence to which the sample was irradiated

\DeclareSIUnit{\neq}{n_{eq}} % neutron equivalent
\DeclareSIUnit{\electron}{e^-} % electron
\DeclareSIUnit{\sample}{Sa} % For oscilloscope sampling frequency

\begin{document}
\maketitle
\flushbottom

\section{Introduction}
In recent years the research and development of silicon detectors for tracking
purposes in high energy physics experiments has shifted to include not only
requirements for high spatial resolution and sufficient radiation hardness, but
also accurate time resolution~\cite{Sadrozinski_2017_4D_particle_tracking}. With
the forthcoming upgrades of the LHC to higher luminosities
(HL-LHC~\cite{Aberle_2020_TDR_HL-LHC}) and plans for the
FCC~\cite{Abada_2019_FCC}, accurate temporal information will play an important
role in tracking, as spatial information alone is insufficient for adequate
performance at the expected increased spatial density of collisions. For
example, during the Phase-II upgrade of the ATLAS
experiment~\cite{ATLAS_2012_PhaseII_upgrade_LOI}, a new High Granularity Timing
Detector will be installed to provide track time information with a resolution
below \SI{50}{\ps}~\cite{ATLAS_2020_HGTD_TDR}, enhancing the performance of the
planned new tracking system. Various studies of time resolution dealing with
different silicon detector technologies have already been carried out, see for
example~\cite{Agapopoulou_2022_HGTD_LGAD_beam_tests},
\cite{Betancourt_2022_3D_detector_timeResolution}
and~\cite{Iacobucci_2022_BiCMOSHexagonalPrototype_timeResolution}.

In this paper, measurements of time resolution of a Depleted Monolithic Active
Pixel Sensor~(DMAPS) prototype are presented. As a promising alternative to the
well-established hybrid silicon detector, DMAPS technology features a
single-wafer design with readout electronics integrated directly onto the
sensing chip~\cite{Peric_2007_DMAPS}. This configuration provides potential
benefits including lower cost, shorter time of production and a lower material
budget within the tracking volume, all important aspects for future trackers
with large sensor area requirements. DMAPS technology is therefore being
considered for use in future high-rate and high-radiation environments that such
trackers would face.

Within the studies of the prototype, time resolution of a monolithic sensor was
measured for the first time using focused pulsed laser light in Back- and
Edge-TCT~\cite{Kramberger_2010_EdgeTCT} configurations, allowing for
position-sensitive measurements of the time resolution within the depleted
region. Measurements were also performed with a setup using electrons from a
radioactive $\Sr$ source. In all cases the time resolution was determined as a
function of the amount of charge collected on the input of the readout
electronics.

\section{RD50-MPW2 prototype}
\begin{figure}
	\centering
	\includegraphics[width=9cm]{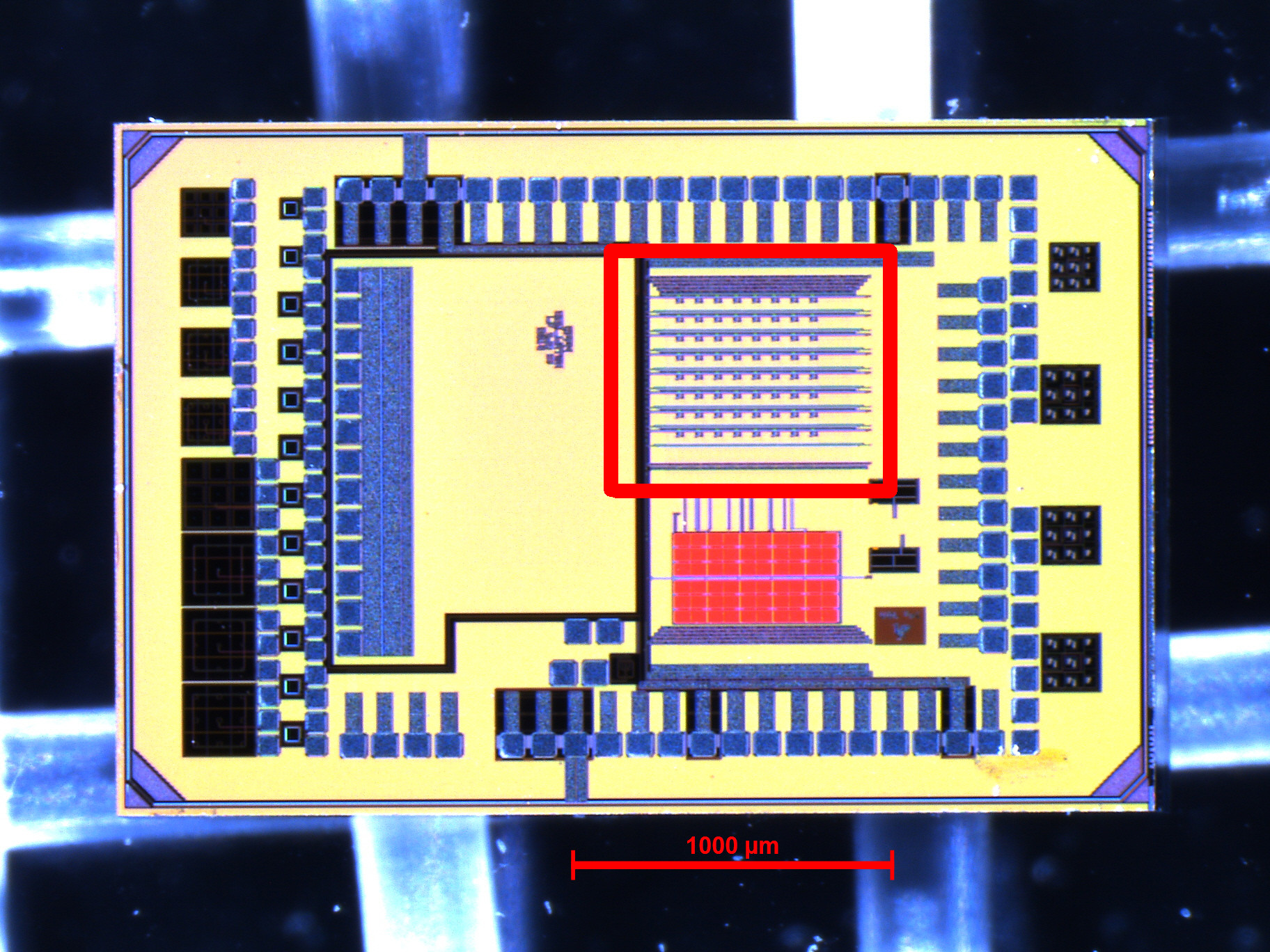}
	\caption[Photograph of RD50-MPW2 under a microscope]{
		Photograph of the RD50-MPW2 prototype under a microscope. The $8\times8$
		active pixel matrix is encircled with a red square.
	}
	\label{fig:MPW2}
\end{figure}

Characterization of time resolution was performed with a prototype detector
named RD50-MPW2 produced by the CERN-RD50
collaboration~\cite{Zhang_2020_RD50MPW2} that is shown in figure~\ref{fig:MPW2}.
It is the second prototype in the series of HV-CMOS monolithic detectors that
the collaboration is developing with the goal of studying and improving the
technology for future tracking applications in particle physics
experiments~\cite{Vilella_2022_RD50MPW_developments}. The prototype is
manufactured in a \SI{150}{\nm} HV-CMOS process from LFoundry on \SI{280}{\um}
thick p-type substrate silicon wafers with varying initial resistivities;
samples with an initial resistivity of \SI{1.9}{\kilo\ohm\cm} were selected for
measurements in this work. In order to study potential effects of radiation
damage on the time resolution, a sample irradiated with reactor neutrons was
measured alongside unirradiated ones. Irradiations were conducted with neutrons
to a \SI{1}{\MeV} neutron equivalent fluence of \fluence with the TRIGA nuclear
reactor at Jožef Stefan Institute~(JSI) in Ljubljana~\cite{Snoj_2012_TRIGA,
Ambrozic_2017_TRIGA_dose}. During neutron irradiation, samples were also exposed
to a total ionizing dose of about \SI{5}{\kilo\gray}.

\begin{figure}
	\centering
	\includegraphics[width=11cm]{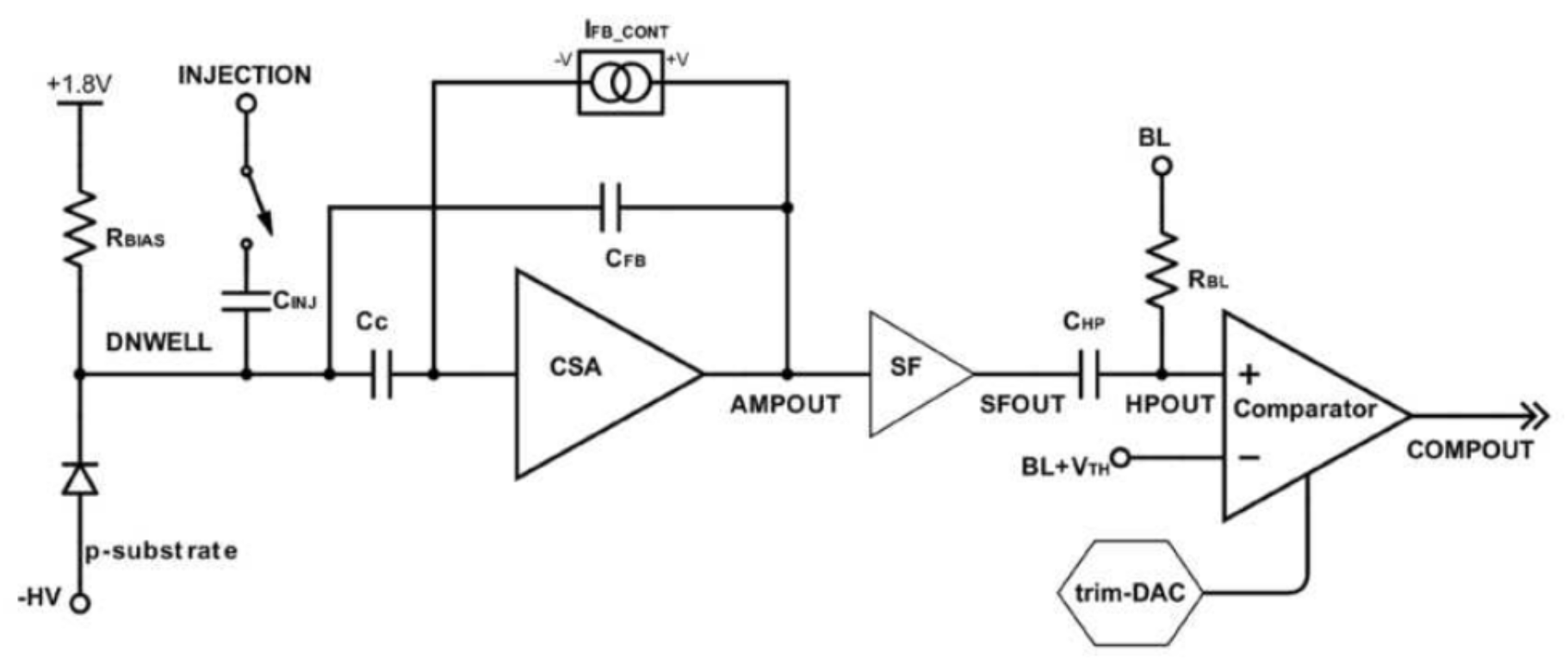}
	\caption[Continuous reset pixel circuit diagram]{
		Circuit diagram of the continuous reset pixel readout electronics.
		(Figure from ref.~\cite{MarcoHernandez_2021_DMAPS_developments_at_RD50},
		\href{https://creativecommons.org/licenses/by/4.0/}{CC~BY~4.0})
	}
	\label{fig:cont_reset_pixel}
\end{figure}

Measurements were made with the $8\times8$ matrix of active pixels, marked with
a red square in figure~\ref{fig:MPW2}. Each pixel in the matrix has a size of
$\SI{60}{\um}\times\SI{60}{\um}$ and contains an analog readout circuit. The
circuit consists of a Charge Sensitive Amplifier~(CSA) and a comparator with a
4-bit trim-DAC for correcting threshold variations arising from manufacturing
nonuniformities. Two types of pixels are implemented in the matrix, differing by
the way the CSA is reset. For most measurements in this study, the so-called
continuous reset pixel (columns \numrange{0}{3} in the matrix), shown in
figure~\ref{fig:cont_reset_pixel}, was chosen due to its output signal Time Over
Threshold~($\mathrm{ToT}$) being linearly proportional to the collected charge.
This is achieved by the constant current source linearly discharging the
feedback capacitor storing the collected charge. In the other type of pixel,
called the switched reset pixel (columns \numrange{4}{7}), the feedback
capacitor is discharged via a much larger current controlled by the comparator
output. Additionally, all pixels chosen for time resolution measurements were
required to have neighboring pixels on all sides to avoid any edge effects.

Configuration and biasing of the chip is implemented via the CaRIBOu
system~\cite{Liu_2017_Caribou, Vanat_2020_Caribou}, which enables setting
various voltages and DAC values of the pixel matrix. For the studies presented
here, a comparator baseline of \SI{900}{\mV} was used at two threshold voltages
of \SIlist{950;1000}{\mV}, corresponding to a threshold in electrons of
\SIlist{1.2;2}{\kilo\electron} respectively for trim-DACs set to
\num{0}~\cite{Hiti_2021_MPW2_timeWalk}. Considering a breakdown voltage of
passive structures at around
\SI{120}{\V}~\cite{Mandic_2022_MPW_passive_pixel_results}, the bias voltage was
set to \SI{100}{\V} for all measurements.

Time walk properties of the prototype were already characterized
in~\cite{Hiti_2021_MPW2_timeWalk}, with measurements showing a good in-time
efficiency (delays below \SI{25}{\ns}) across the entire range of collected
charge and no significant loss of performance after neutron irradiation. In
terms of characterizing the RD50-MPW2 performance, this paper augments the
aforementioned results and presents an in-depth study of the time resolution.

\subsection{Output signal calibration}
\label{sec:calibration}
Each pixel within the matrix contains a calibration circuit enabling the
injection of charge into the front-end electronics. Charge is injected by
connecting a voltage step function via an injection capacitance with a value of
$C_\mathrm{inj}=\SI{2.8}{\fF}$. The amount of injected charge can then be
determined as $e=C_\mathrm{inj}U_\mathrm{inj}$, where $U_\mathrm{inj}$ is the
amplitude of the voltage step function.

\begin{figure}
	\centering
	\includegraphics[width=10cm]{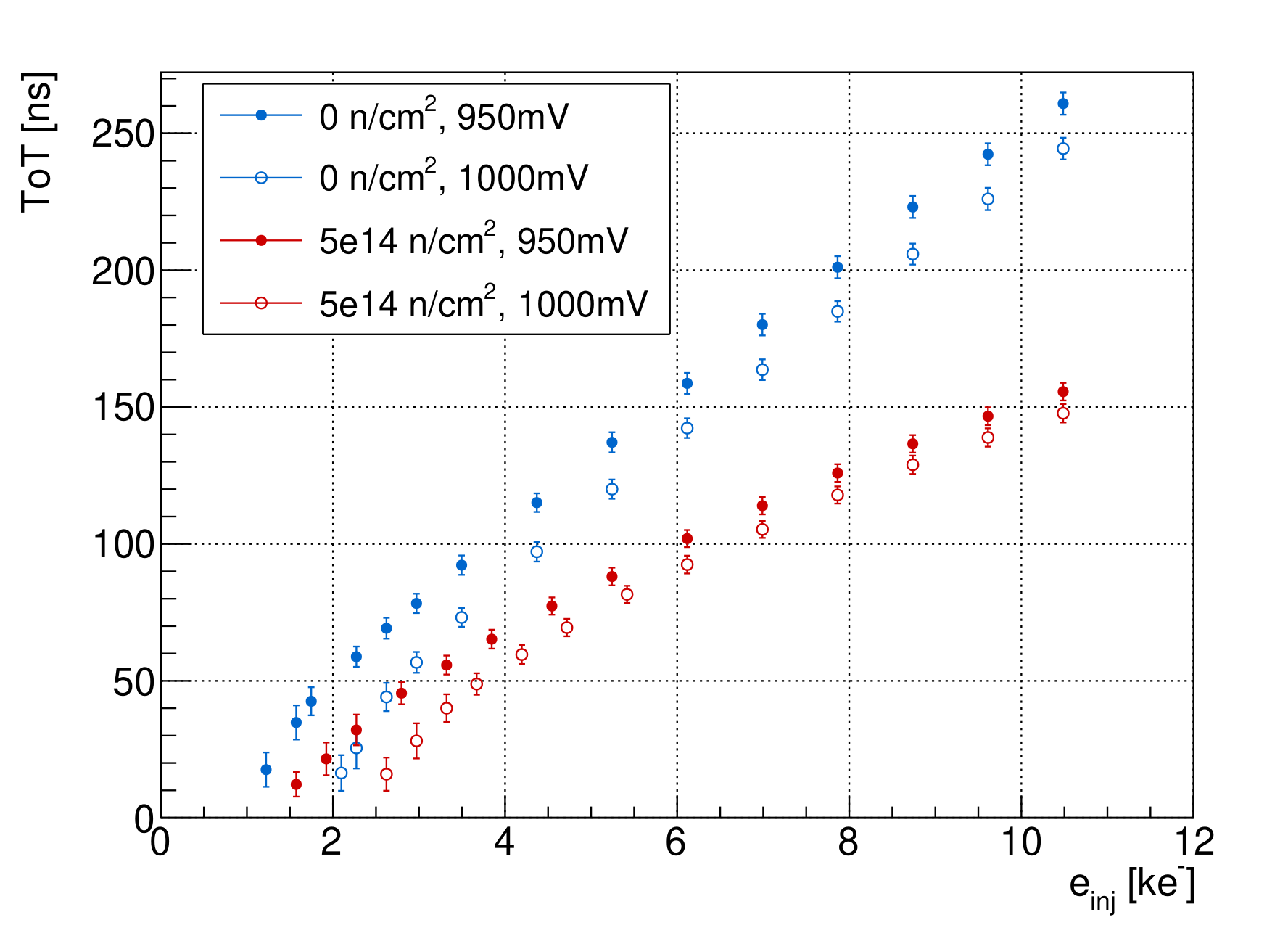}
	\caption[Output signal $\mathrm{ToT}$ calibration]{
		Example output signal $\mathrm{ToT}$ calibration curves for two
		different pixels operated at two threshold voltages. One pixel (blue)
		was measured before irradiation, the other (red) after irradiation to
		\fluence. Error bars show the standard deviation in \num{1000} events
		measured at each charge value.
	}
	\label{fig:ToT_calibration}
\end{figure}

By varying $U_\mathrm{inj}$, a calibration of the comparator $\mathrm{ToT}$ to
the amount of injected charge can be performed. For example,
figure~\ref{fig:ToT_calibration} shows calibration curves for pixels later
selected for time resolution measurements with the Edge-TCT method. The data
points show a large variation of $\mathrm{ToT}$ values between the two pixels.
However, by measuring multiple pixels across the matrix and comparing their
outputs before and after irradiation, it was shown that irradiation does not
affect the outputs significantly~\cite{Hiti_2021_MPW2_timeWalk}. The differences
seen in figure~\ref{fig:ToT_calibration} are therefore mostly a result of
deviations in the CSA gain (feedback capacitance) arising during the production
of samples. Consequently, individual calibration of each pixel measured is
necessary for correct determination of collected charge.

\section{Time resolution measurements}
Measurements of time resolution were performed with the Transient Current
Technique~(TCT) using a pulsed and focused laser beam with a $\mathrm{FWHM}$ of
less than \SI{6}{\um} at the focusing point. By focusing the incident laser
light, free charge carriers can be created at different positions within the
depleted region of the pixel, thus enabling determination of time resolution as
a function of the position where the charge was created. To perform these
position-sensitive measurements, the carrier board is placed onto precision
stages with a movement accuracy below \SI{1}{\um}. The time resolution was
measured using two different setups varying by the incident laser beam
orientation.

\begin{figure}
	\centering
	\includegraphics[width=0.49\textwidth]{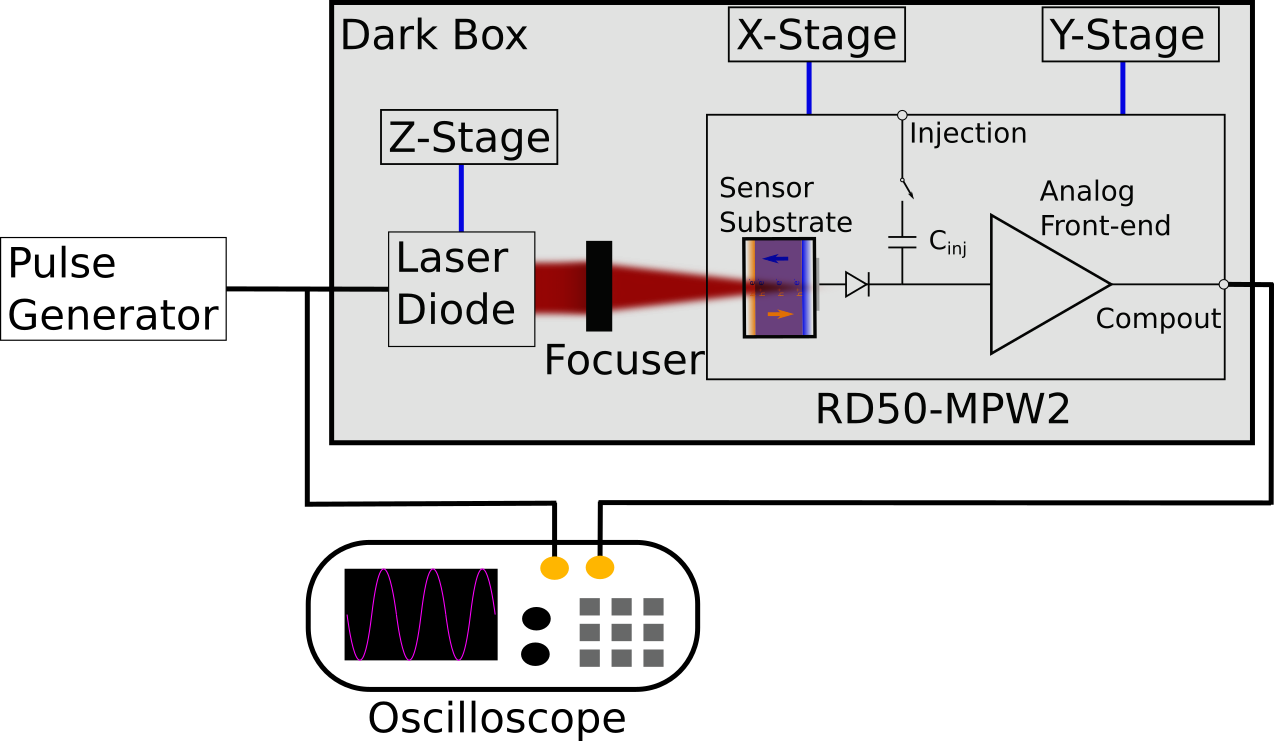}
	~
	\includegraphics[width=0.49\textwidth]{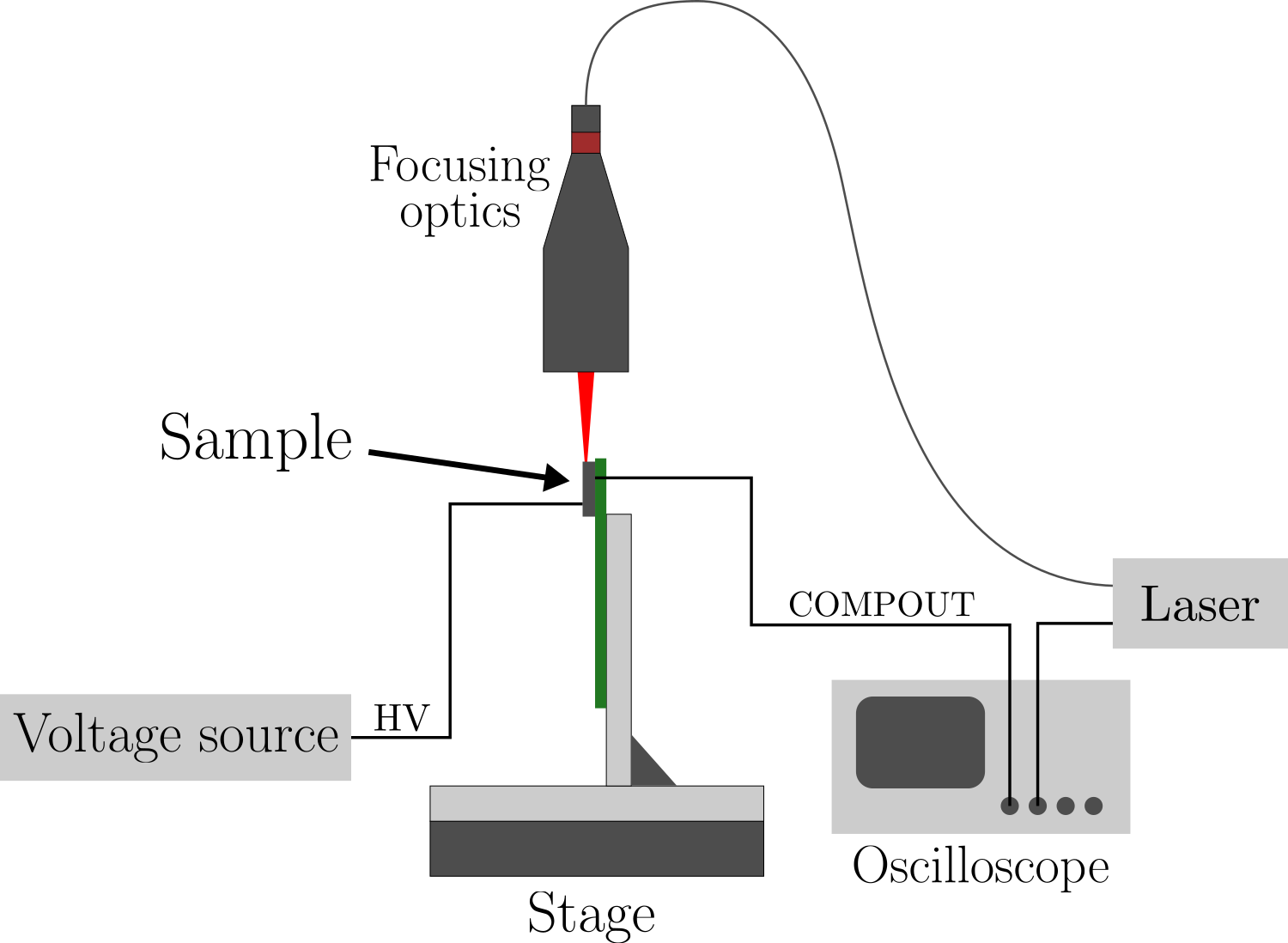}
	\caption[Schematic depiction of the TCT setups]{
		Schematic depiction of the Backside-TCT setup at Nikhef~(left) and
		Edge-TCT setup at JSI~(right).
	}
	\label{fig:TCT_setups}
\end{figure}

Backside-TCT measurements were performed with a \SI{980}{\nm} laser setup
located at Nikhef. A schematic of the setup is shown on the left in
\cref{fig:TCT_setups}. The light is injected into the sensor backside through a
hole in the carrier PCB. The laser is driven via a pulse generator that was kept
on a rise and fall time of \SI{2}{\ns}, a pulse width of \SI{10}{\ns}, and an
amplitude of \SI{2.5}{\V}. The laser intensity was adjusted via a variable
optical attenuator controlled via a power supply. Both the comparator output and
the pulse generator signal are recorded using a DSO3000 series oscilloscope with
a \SI{500}{\MHz} bandwidth and a \SI{4}{\giga\sample\per\s} sampling rate.

Edge-TCT measurements were performed at the JSI laboratories using a modified
version of the Particulars setup~\cite{Particulars_homePage} shown on the right
in figure~\ref{fig:TCT_setups} (more details on the setup can be found
in~\cite{Hiti_2021_MPW2_timeWalk}). In this setup, \SI{1064}{\nm} laser light
with a pulse duration of \SI{300}{\ps} was oriented to enter the sample from the
side edge of the sample. Signals were sampled with a DRS4 oscilloscope with an
analog bandwidth of \SI{700}{\MHz} and sampling rate of
\SI{3.5}{\giga\sample\per\s}.

\begin{figure}
	\centering
	\includegraphics[width=8.5cm]{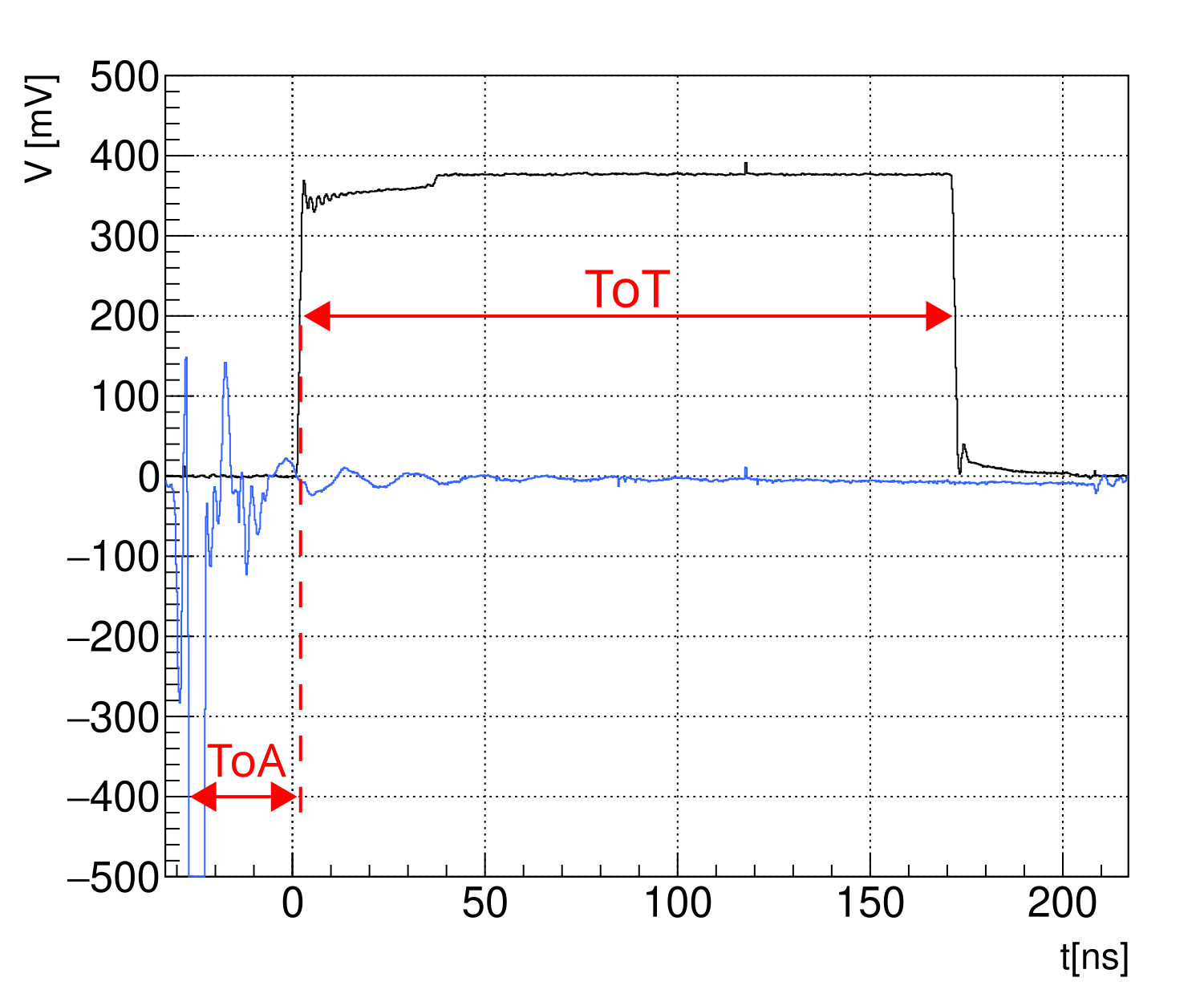}
	\caption[Example event with the Edge-TCT setup]{
		Example event as acquired using the Edge-TCT setup. The positive signal
		(in black) is the output of the in-pixel comparator. The negative signal
		(blue) is the trigger signal from the laser driver. Thresholds for level
		crossing times were chosen here at \SIlist{200;-400}{\mV} for the pixel
		output and laser signal respectively. Time resolution is determined from
		the spread of $\mathrm{ToA}$.
	}
	\label{fig:ETCT_eventExample}
\end{figure}

A signal from the pulse generator (laser driver), which generates the laser
pulses, was used as a time reference for determining the arrival time of the
pixel response. Figure~\ref{fig:ETCT_eventExample} shows a typical event. The
time of arrival~($\mathrm{ToA}$) of the pixel response relative to the reference
signal is determined as the difference of level crossing times over a fixed
threshold. The $\mathrm{ToT}$ of the pixel output is also recorded and used for
determining the amount of collected charge using the calibration curves
described in section~\ref{sec:calibration}. The time resolution is obtained as
the standard deviation of the arrival times~($\sigma_\mathrm{ToA}$) over
multiple collected events. $\mathrm{ToA}$ is only determined up to an additive
constant, which is affected by optical fiber and cable lengths. This, however,
does not affect the time resolution calculation, since only the relative spread
of these times is of interest.

The total time resolution of a detector is a result of various
contributions~\cite{Sadrozinski_2017_4D_particle_tracking} and can be written as
their sum
\begin{equation}
	\sigma_t^2=\sigma_\mathrm{Lan}^2+\sigma_\mathrm{dis}^2+\sigma_\mathrm{tw}^2+
	\sigma_\mathrm{j}^2\,,
\end{equation}
where $\sigma_\mathrm{Lan}$ is the effect of Landau fluctuations and
$\sigma_\mathrm{dis}$ the effect of signal distortion due to non-uniformities in
the weighting field and charge carrier drift velocities. The contribution from
time walk $\sigma_\mathrm{tw}$ is not relevant in this case since the time
resolution was measured as a function of collected charge (i.e.\ at constant
signal pulse amplitudes). The jitter of the electronics $\sigma_\mathrm{j}$ is
given by the expression
$\sigma_\mathrm{j}=t_\mathrm{r}/(S/N)$~\cite{Cartiglia_2014_UFSDperformance},
where $t_\mathrm{r}$ is the signal rise time and $S/N$ the signal-to-noise
ratio, giving a $1/S$ dependence on the collected charge.

\subsection{Time resolution with Backside-TCT}
For measurements of time resolution with Backside-TCT, two types of measurements were performed on an unirradiated sample using the laser setup: a scan over the entire $8\times 8$ pixel matrix and an in-pixel position-sensitive scan. For all measurements the laser was focused on the pixel center\footnote{The center of a pixel is determined via a simple scan in $x$ and $y$ by taking the midpoint between the edges corresponding to a drop in the pixel response.} and moved along the row and column direction to perform the full matrix and in-pixel scan.
%The center of a pixel is determined via a simple x- and y-line scan looking at the pixel response determining the center as the middle between the edges corresponding to a drop of the pixel response. For the full matrix scan, the laser was aligned in the center of pixel (2,2) and the stepping of the motor stage was aligned such that a 60 micron movement along the row and column axes, which was verified using pixel (6,6) is performed. This movement begins at (0,0) and moves to (7,0) after which the movement is repeated starting at (0,1) until pixel (7,7) is reached.
 %Furthermore, an in-pixel scan was performed in which the laser is centered on a pixel after which it is moved to $\vec{p}_{\mathrm{stop}} = [-60, -60]\, \mu \mathrm{m}$ along the column and row axis. The laser then moves by \SI{3}{\micro m} steps along the row to $\vec{p}_{\mathrm{stop}} = [60, -60]\, \mu \mathrm{m}$. Afterwards the laser is returned to the starting position of the current row movement, moves up \SI{3}{\micro m} to $\vec{p}_{\mathrm{stop}} = [-57, -60]\, \mu \mathrm{m}$ and repeats the measurement until the laser has moved to position $\vec{p}_{\mathrm{stop}} = [60, 60]\, \mu \mathrm{m}$.
Each measurement step includes at least 50 measured waveforms triggered on the pulse generator and measurement steps with fewer than 10 total responses of the comparator output registered in the waveforms are discarded.

\subsubsection{Full matrix}
All measurements of the full matrix are conducted with trim-DAC optimized settings and the comparator threshold set to \SI{1000}{\mV}. Taking the trim-DAC adjustments into account, the effective thresholds of the two pixel flavors at trim-optimized settings differ from one another, with switched reset pixels showing an effective threshold of \SI{1460}{\electron} while continuous reset pixels show a threshold of \SI{2980}{\electron}~\cite{tsolantaMPW2}.

\begin{figure}
	\centering
	\includegraphics[width=0.4\textwidth]{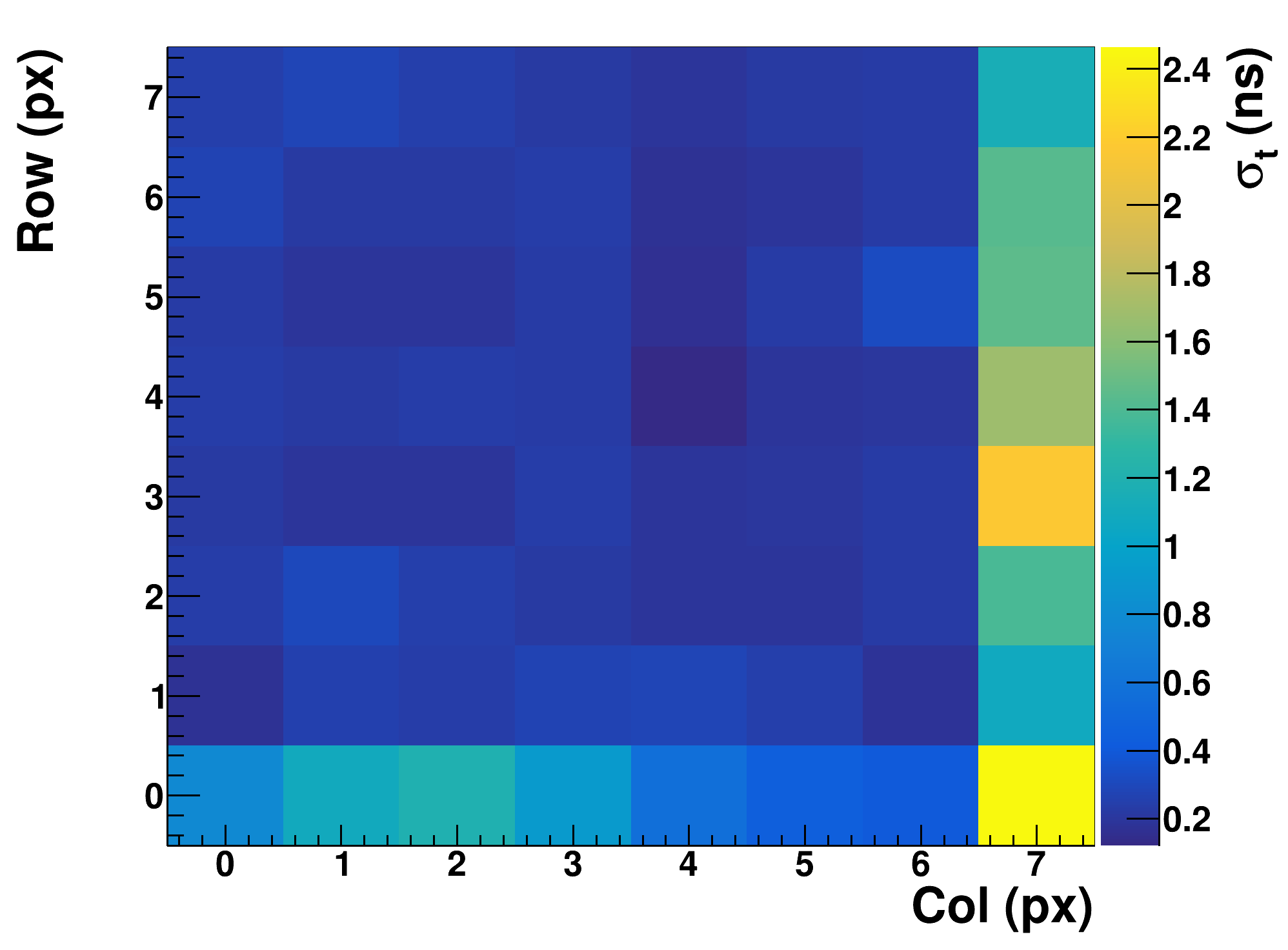}
	\qquad
	\includegraphics[width=0.4\textwidth]{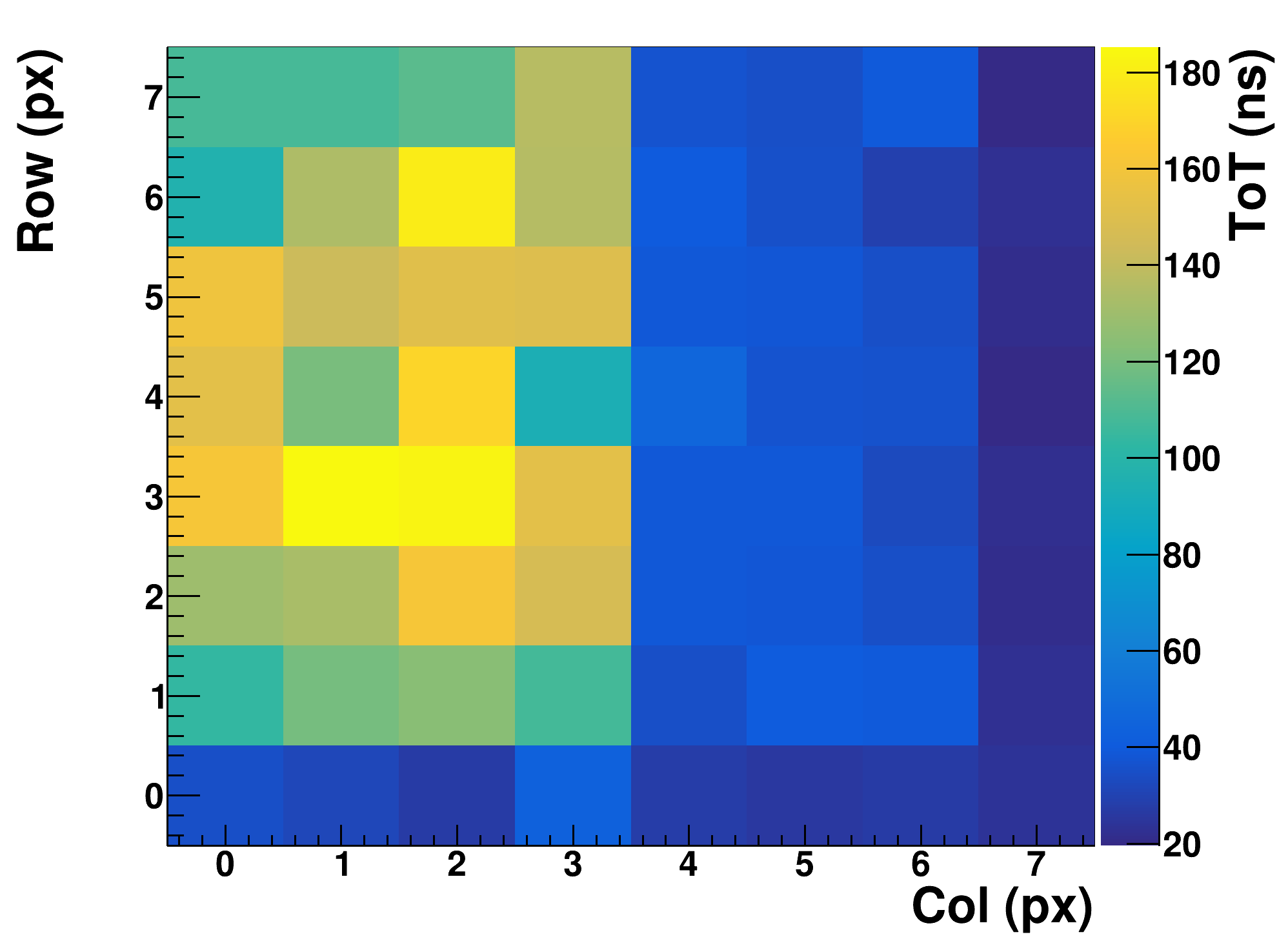}
	\caption[Full matrix time resolution and $\mathrm{ToT}$]{
		Time resolution (left) and measured $\mathrm{ToT}$ (right) of the
		RD50-MPW2 pixel matrix at about \SI{12100}{\electron} equivalent charge
		injection via laser centered on each pixel.
	}
	\label{fig:las_time_res_n_ToT_2D}
\end{figure}

The left side of \cref{fig:las_time_res_n_ToT_2D} shows the time resolution achieved for the full pixel matrix split into their row and column ID at a laser induced charge injection value of about \SI{12}{\kilo\electron}. Overall the response of the pixel matrix is uniform, showing a time resolution of about \SI{210}{\ps} for all pixels with the exception of row 0 and column 7 which show a worse time resolution.
A look at the measured $\mathrm{ToT}$ of each pixel, depicted in the right plot of \cref{fig:las_time_res_n_ToT_2D}, shows that the measured $\mathrm{ToT}$ in row 0 and column 7 are far lower relative to the response measured by the other pixels. This discrepancy was also present upon repeat of the measurement.
%A look at the measured charge of each pixel, depicted in the right plot of \cref{fig:las_time_res_n_ToT_2D}, shows that the measured charge in row 0 and column 7 are lower relative to the response measured by the other pixels. This discrepancy was also present upon repeat of the measurement.
A further investigation with an in-pixel measurement showed that both the amount and location of maximum induced charge value for pixel $(0,0)$\footnote{Pixel locations are given as $(\mathrm{col}, \mathrm{row})$ values.} is further than \SI{60}{\um} away along the column from the center of pixel $(0,1)$.
This is most likely due to the electric field of the pixels at the edge of the matrix not being constrained by surrounding structures resulting in a non-uniform response of the pixel. The response differs for the four edges, as the structures surrounding the $8\times 8$ matrix also differ from one another on all sides.
This was confirmed with an in-pixel measurement of $(0,0)$ which gathered charge from larger distances than expected while having a \SI{30}{\percent} lower charge response in the pixel center than pixels located in the central $6\times 6$ matrix.
As a result, in all further measurements only the central $6\times 6$ matrix is shown to avoid these boundary effects.

Another effect visible in \cref{fig:las_time_res_n_ToT_2D} is a far lower $\mathrm{ToT}$ for the right half of the matrix. This is due to the different pixel flavor as the switched reset pixels are drained far quicker once the signal reaches the threshold~\cite{Vilella_2022_RD50MPW_developments}. As such the linearity between charge and $\mathrm{ToT}$ is not fulfilled. This has no impact on the time resolution but all values of charge refer to the results given by the continuous reset pixels after $\mathrm{ToT}$ to charge conversion from the calibration for which the linearity is true.

\begin{figure}
	\centering
	\includegraphics[width=0.65\textwidth]{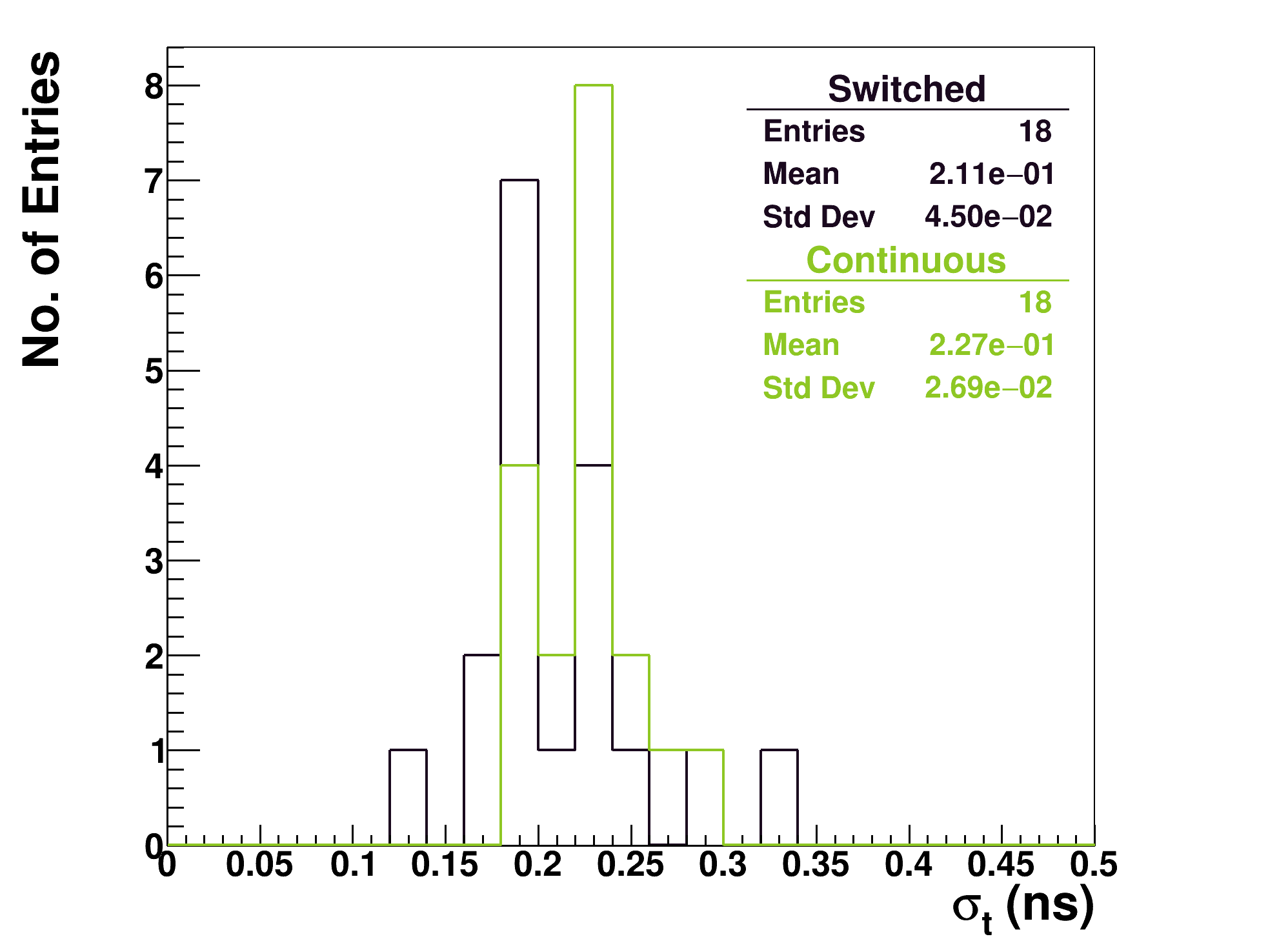}
	\caption[Distributions of pixel time resolutions]{
		Comparison between the time resolution achieved via minimum ionizing
		particle equivalent charge induction from a \SI{980}{\nm} laser  for
		switched pixel flavor with those of the continuous reset pixel flavor.
	}
	\label{fig:time_res_pixel_flavor_compare}
\end{figure}

The core of the matrix shows good agreement between the time resolution of the switched reset pixels and the continuous reset pixels, see \cref{fig:time_res_pixel_flavor_compare}. The switched reset pixels show a mean time resolution of $\sigma_{t,\mathrm{switch}} = \SI{211\pm 45}{\ps}$ compared to the continuous reset pixels which have a time resolution of $\sigma_{t,\mathrm{cont}} = \SI{227\pm 27}{\ps}$. Though there is a small difference, the two pixel flavors time resolutions are still within the error of one another.
The measurements are also congruent with the results achieved through direct charge injection only probing the front-end which is on the order of $\sigma_{t,\mathrm{switch\ injection}} = \SI{187\pm 24}{\ps}$ and $\sigma_{t,\mathrm{cont. injection}} = \SI{202\pm 22}{\ps}$; the slightly worse performance of the continuous reset pixels results from different thresholds for the two types of pixels.

\subsubsection{In-pixel}
Multiple in-pixel measurements were performed for different pixel flavors. Shown here are the results for the continuous reset pixel $(3,2)$ which is located in the inner matrix core on the boundary changing the pixel flavor to the switched reset pixels.

\begin{figure}
	\centering
	\includegraphics[width=0.65\textwidth]{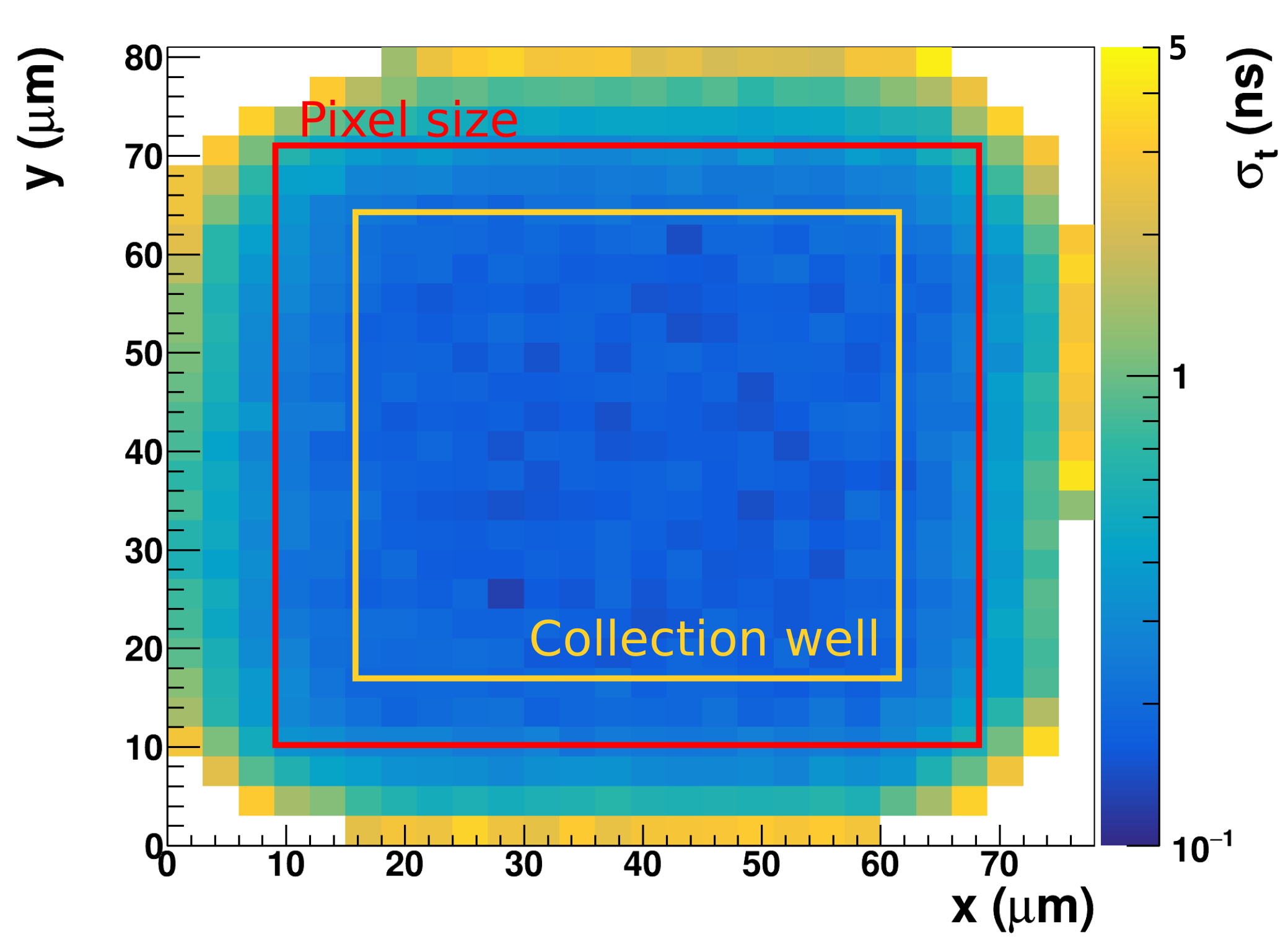}
	\caption[In-pixel scan of time resolution]{
		In-pixel scan of the time resolution of pixel $(3,2)$ through charge
		injection via laser of about \SI{13400}{\electron}. The physical
		location of the collection well (orange square) and the pixel boundary
		(red square) are drawn into the picture. Positions with insufficient
		measurements due to insignificant amount of hits above threshold are
		shown in white.
	}
	\label{fig:in-pix_time-res_3-2}
\end{figure}

The charge injected via laser was slightly above the value injected via the full matrix scan. The time resolution achieved via the in-pixel scan is depicted in \cref{fig:in-pix_time-res_3-2}. The red square corresponds to the $\SI{60}{\um}\times \SI{60}{\um}$ pixel boundary while the yellow square corresponds to the $\SI{46}{\um}\times \SI{46}{\um}$ collection well. The area within the pixel boundary shows a flat time resolution of about $\sigma_{t,\mathrm{well}} = \SI{173\pm 15}{\ps}$ beneath the collection well and a time resolution of $\sigma_{t,\mathrm{pixel}} = \SI{188\pm 32}{\ps}$ within the pixel boundary.
The slightly improved resolution is due to the aforementioned slightly higher charge injection value of $\approx \SI{13400}{\electron}$. The results show some charge sharing up to a distance of $\approx\SI{10}{\um}$ with the given statistics.

\begin{figure}
	\centering
	\includegraphics[width=0.45\textwidth]{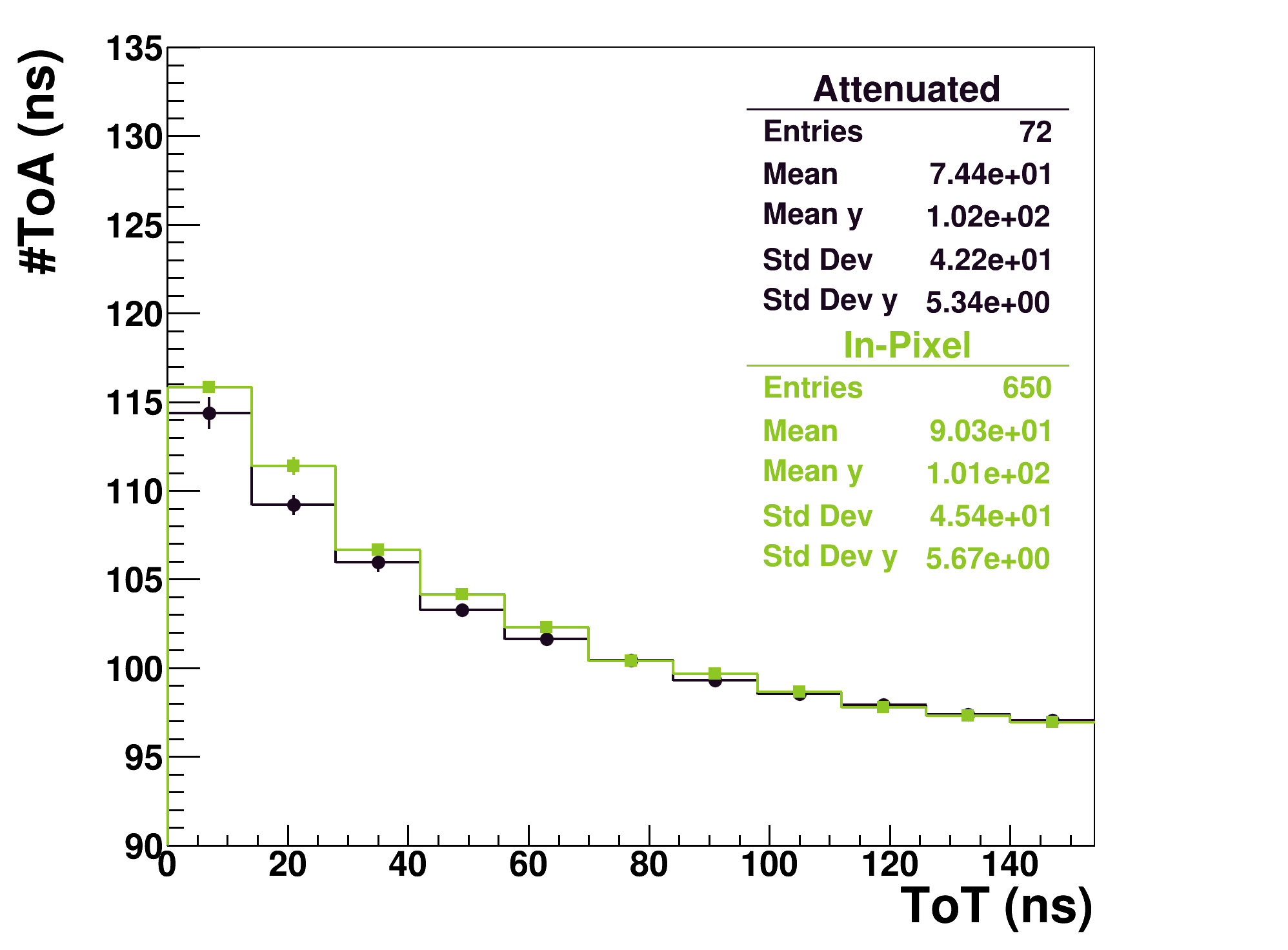}
	\includegraphics[width=0.45\textwidth]{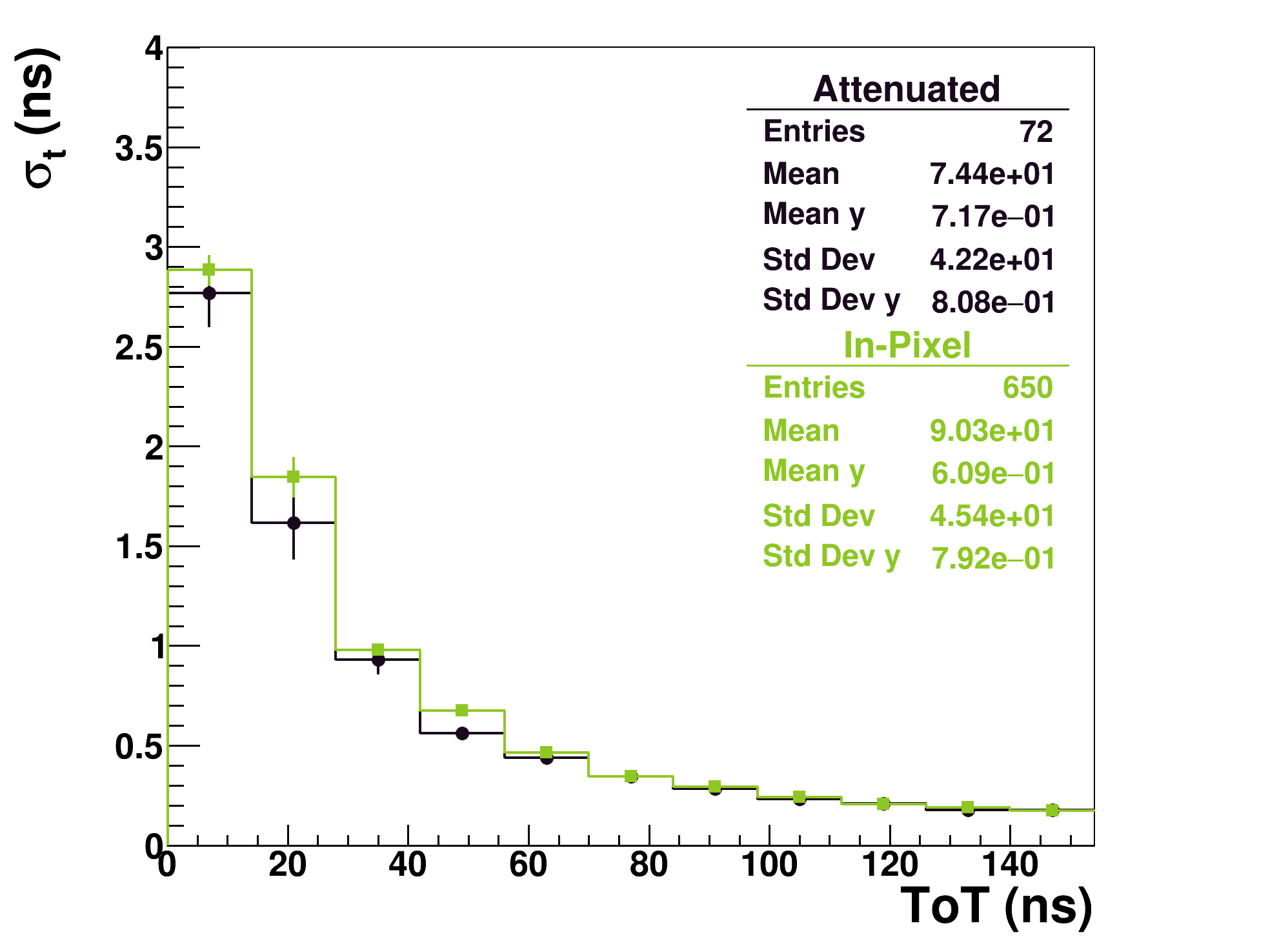}
	\caption[
		$\mathrm{ToA}$ and time resolution comparison for attenuated and
		in-pixel scan
	]{
		Comparison of the $\mathrm{ToA}$ distribution relative to the measured
		charge in $\mathrm{ToT}$ (left) and the measured time resolution
		relative to the measured charge in $\mathrm{ToT}$ (right) for pixel
		$(3,2)$. The black plot corresponds to the center-of-pixel measurement
		with an attenuated laser pulse amplitude, whereas results shown in green
		are obtained at constant laser pulse amplitude by scanning the entire
		pixel area.
	}
	\label{fig:attenuated_in_pixel_compare}
\end{figure}

Beyond the pixel boundary, the time resolution begins to drop rapidly. However, further investigation is needed to determine whether the increased $\mathrm{ToA}$ and $\sigma_t$ at the edge are purely due to increased time walk as a result of lower charge, or whether the geometric distance to the collection well adds an additional contribution.
For this purpose, the measured $\mathrm{ToA}$ and $\sigma_t$ gathered via the in-pixel measurements are compared with a measurement in which the laser was kept focused on the pixel center and the injected charge was adjusted via the optical attenuator.
These comparisons are depicted in \cref{fig:attenuated_in_pixel_compare} for the $\mathrm{ToA}$ (left) and the $\sigma_t$ (right). Both distributions show excellent overlap at high $\mathrm{ToT}$ values at which point the in-pixel scan is also located over the center of the pixel. However, at $\mathrm{ToT}$ values below \SI{80}{\ns} the measured $\mathrm{ToA}$ and $\sigma_t$ from in-pixel measurements begin to rise faster than the results gathered via the centered attenuated signal.
At low $\mathrm{ToT}$, measurements are performed close to the threshold which also increases the statistical uncertainty of the measurements. Nonetheless, at very low charge values, the expected $\mathrm{ToA}$ differs by up to \SI{3}{\ns} on average while the time resolution is worse by \SI{150}{\ps} indicating a contribution from fluctuations due to inhomogeneous charge collection times.
While not too relevant for laser measurements focused on the pixel center, measurements with beam particles or radioactive sources will be affected by this additional contribution.

\subsection{Time resolution with Edge-TCT}
\begin{figure}
	\centering
	\begin{subfigure}{0.9\textwidth}
		\includegraphics[width=\textwidth]{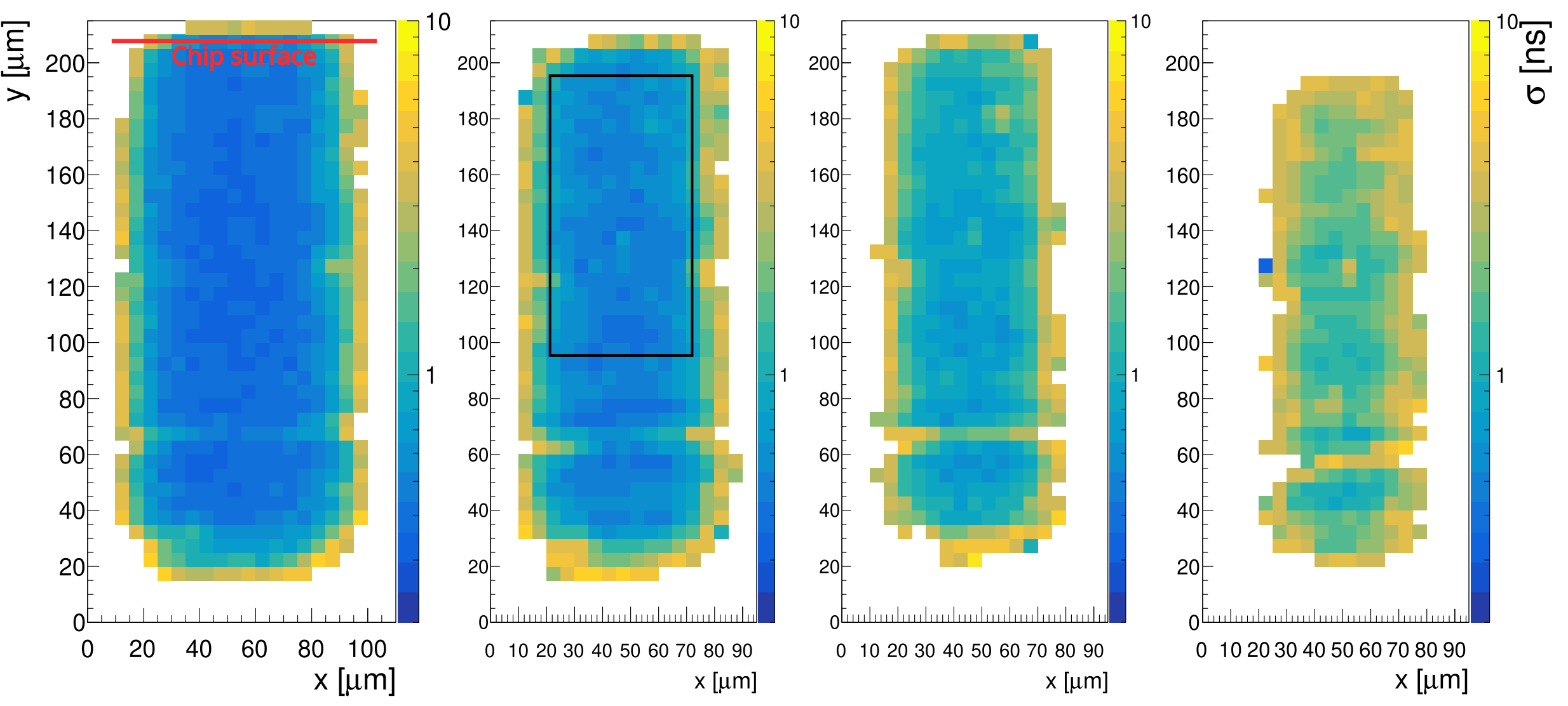}
		\caption{Unirradiated sample}
	\end{subfigure}

	\vspace{5mm}
	\begin{subfigure}{0.9\textwidth}
		\includegraphics[width=\textwidth]{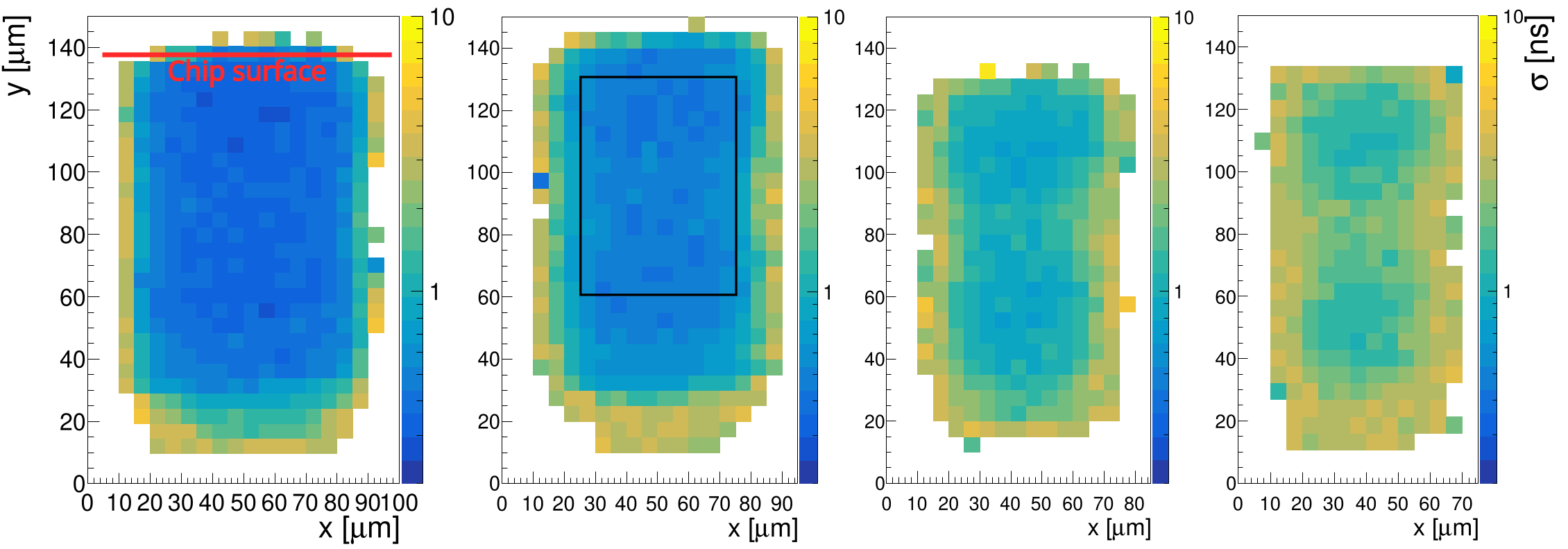}
		\caption{\fluence irradiated sample}
	\end{subfigure}
	\caption[2D measurements of time resolution with Edge-TCT]{
		Two-dimensional measurements of the time resolution with Edge-TCT at
		different laser pulse intensities (pulse intensity is decreasing to the
		right). The chip surface is at the top edge of the plots. The scan area
		was selected to enclose the entire pixel volume, with white areas
		representing locations where no response from the pixel was seen (i.e.\
		locations outside the pixel's active volume). At each point, the time
		resolution is plotted as $\sigma_\mathrm{ToA}$ over \num{100} events.
		Points selected for plots in figure~\ref{fig:ETCT_sigma} are marked by
		the black rectangles.
	}
	\label{fig:ETCT_2D_sigma}
\end{figure}

In the Edge-TCT setup, measurements were performed on single pixels without
trim-DAC optimization (all trim-DACs were set to 0). Two-dimensional profiles of
time resolution for the unirradiated and irradiated samples are presented in
figure~\ref{fig:ETCT_2D_sigma} for different laser beam intensities. The width
of the measured profiles is consistent with the \SI{60}{\um} size of the pixels
with some charge sharing beyond the pixel boundary present, as was seen in the
Backside-TCT measurement. The depletion depth decreases with irradiation due to
the increase of the effective space charge
concentration~\cite{Mandic_2022_MPW_passive_pixel_results}. At the highest laser
intensities, the time resolution reaches a value of around \SI{300}{\ps}, while
at low intensities, it degrades to a value above \SI{1}{\ns}. Time resolution
values are similar throughout the center of the depleted region, indicating
consistent charge collection independent of the initial location of deposition,
while in the charge sharing region on pixel edges, time resolution degrades as
was also seen in results from Backside-TCT. The relatively broad smearing of the
edges is mainly due to a finite laser beam width and suboptimal focusing on
account of the pixel matrix being positioned deep within the chip, possibly
causing beam reflections before the light reaches the pixel. This is also the
likely cause of an irregularity in the unirradiated sample seen at
$y\approx\SI{65}{\um}$.

\begin{figure}
	\centering
	\begin{subfigure}{0.495\textwidth}
		\includegraphics[width=\textwidth]{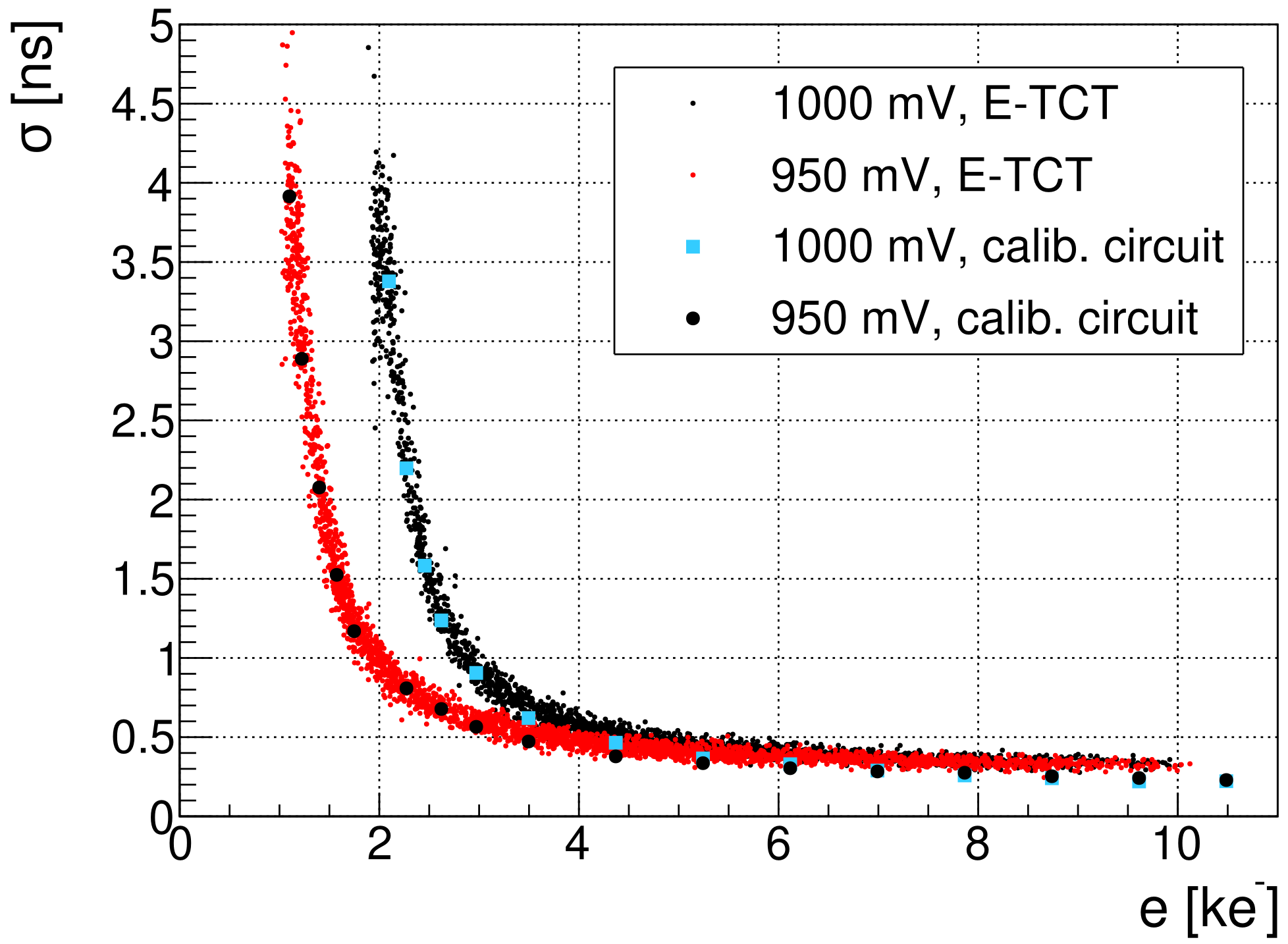}
		\caption{Unirradiated sample}
	\end{subfigure}
	\begin{subfigure}{0.495\textwidth}
		\includegraphics[width=\textwidth]{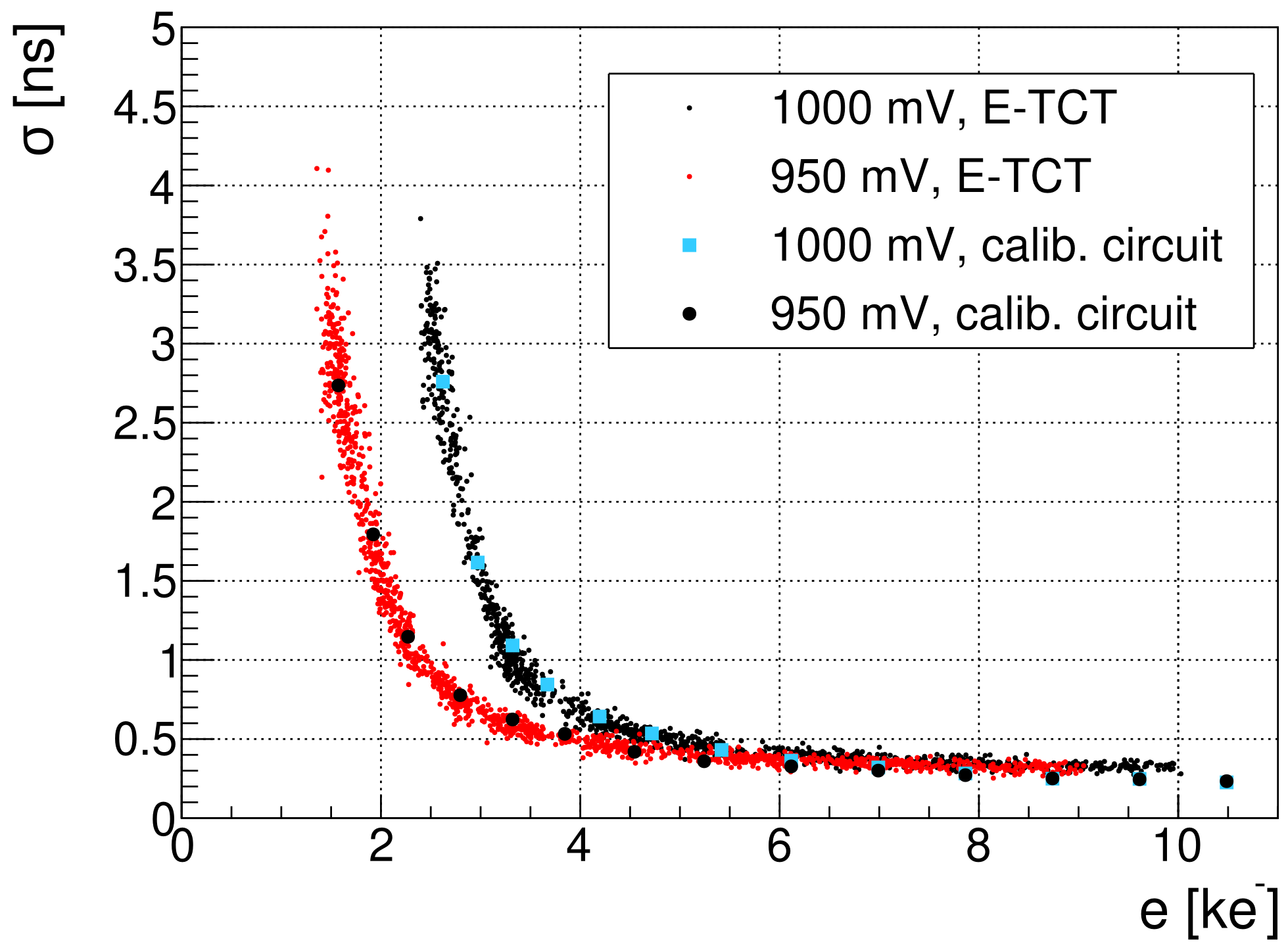}
		\caption{\fluence irradiated sample}
	\end{subfigure}
	\caption[Pixel time resolution vs.\ collected charge for Edge-TCT]{
		Pixel time resolution vs.\ collected charge for Edge-TCT measurements
		for two comparator threshold values. Results are compared with values
		obtained from measurements using the calibration circuit (shown as
		larger sparse points).
	}
	\label{fig:ETCT_sigma}
\end{figure}

To better assess the time resolution, points from the central, most efficient
volume \SI{50}{\um} wide and \SI{100}{\um} (\SI{70}{\um}) in depth for the
unirradiated (irradiated) samples starting \SI{10}{\um} below the chip surface
(regions marked with a black rectangle in figure~\ref{fig:ETCT_2D_sigma}) are
selected and plotted as a function of the collected charge obtained from the
average $\mathrm{ToT}$ value at each point. Two-dimensional scans were taken at
both threshold levels and several laser beam intensities were used to cover the
entire range of collected charge values. Results in figure~\ref{fig:ETCT_sigma}
show a time resolution better than \SI{500}{\ps} for charges above
\SI{5}{\kilo\electron} in all cases and reaching approximately \SI{320}{\ps} at
the highest measured charge of \SI{10}{\kilo\electron}. At low charge, the time
resolution degrades with the point of divergence depending on the comparator
threshold setting. Points of divergence give thresholds in electrons and are
consistent with results obtained by activation curve scans
in~\cite{Hiti_2021_MPW2_timeWalk}, which were done by injecting a variable
amount of charge via the calibration circuit and determining the minimum charge
at which the pixel starts producing an output signal. At the measured depletion
depths, the charge deposited by a MIP has a most probable value of
\SI{14}{\kilo\electron} (\SI{9}{\kilo\electron}) for an unirradiated
(irradiated) sample, which is large enough to lie within the asymptotic part of
the pixel's time resolution dependence and thus provide a good expected timing
performance for MIPs. Irradiation to \fluence does not indicate any significant
degradation of performance; the increase of the point of divergence comes from
the lower CSA gain as discussed in section~\ref{sec:calibration}.

The time resolution was also determined by using the pixel's calibration
circuit. Since the charge is injected directly in front of the readout
electronics, the charge collection in the sensor is not present in this case,
making the jitter of the electronics the main contribution to the time
resolution. Comparing these measurements with the Edge-TCT results, a general
good agreement of values is seen between the two methods, indicating that the
time resolution is dominated by the electronics jitter. A slight increase in the
resolution seen at charges above \SI{6}{\kilo\electron} can be attributed to
other effects from the charge collection phase, or possibly variations in the
intensity of successive laser pulses.

\begin{figure}
	\centering
	\includegraphics[width=0.65\textwidth]{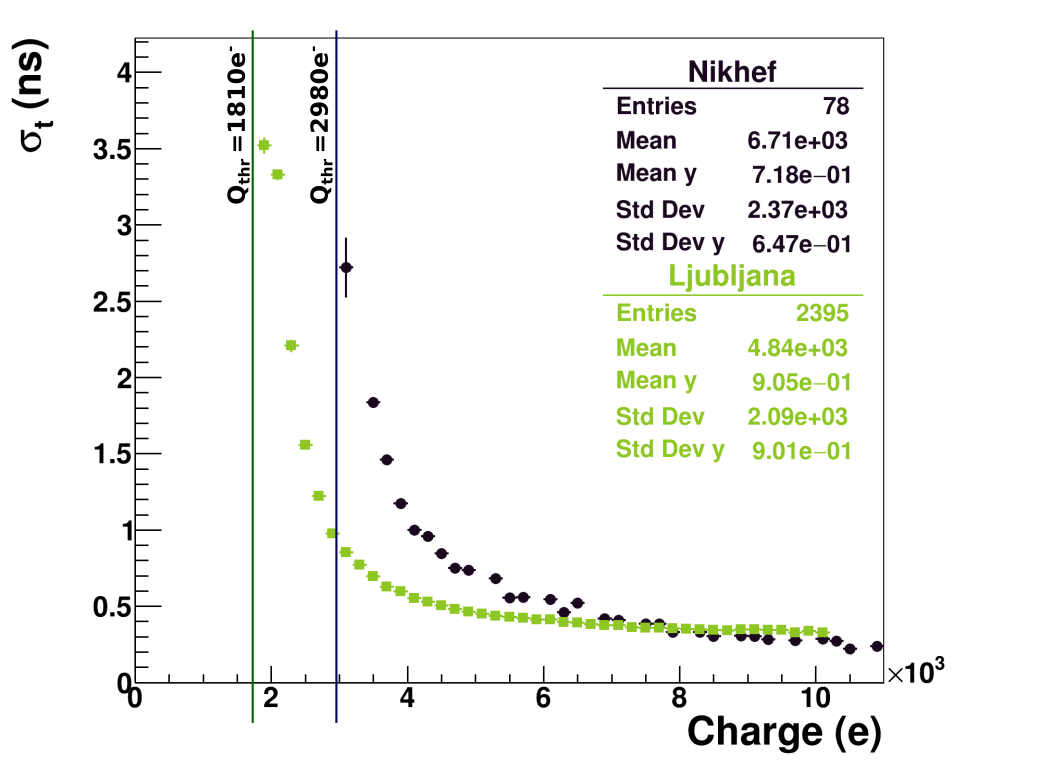}
	\caption[JSI and Nikhef time resolution comparison]{
		Comparison of time resolution obtained via Edge-TCT at JSI (green) with
		the results from Backside-TCT at Nikhef (black), both for a continuous
		reset pixel.
	}
	\label{fig:ljubljana_nikhef_compare}
\end{figure}

A comparison between Edge-TCT measurements performed at JSI with Backside-TCT
measurements from Nikhef for a single pixel are depicted in
\cref{fig:ljubljana_nikhef_compare}. While both measurements use a comparator
threshold of \SI{1000}{\mV}, the trim-DACs at JSI are kept at 0, resulting in a
threshold of \SI{1810}{\electron}. At Nikhef, the trim-DACs are increased in
order to equalize the thresholds between pixels due to the investigation of the
full matrix, resulting in a higher average threshold of \SI{2980}{\electron} for
the continuous reset pixels. Overall, the behavior of the two measurements is in
good agreement showing similar behavior when the laser induced charge is close
to the respective pixel thresholds.

\subsection{Time resolution with \texorpdfstring{$\Sr$}{Sr-90}}
\begin{figure}
	\centering
	\includegraphics[width=6.5cm]{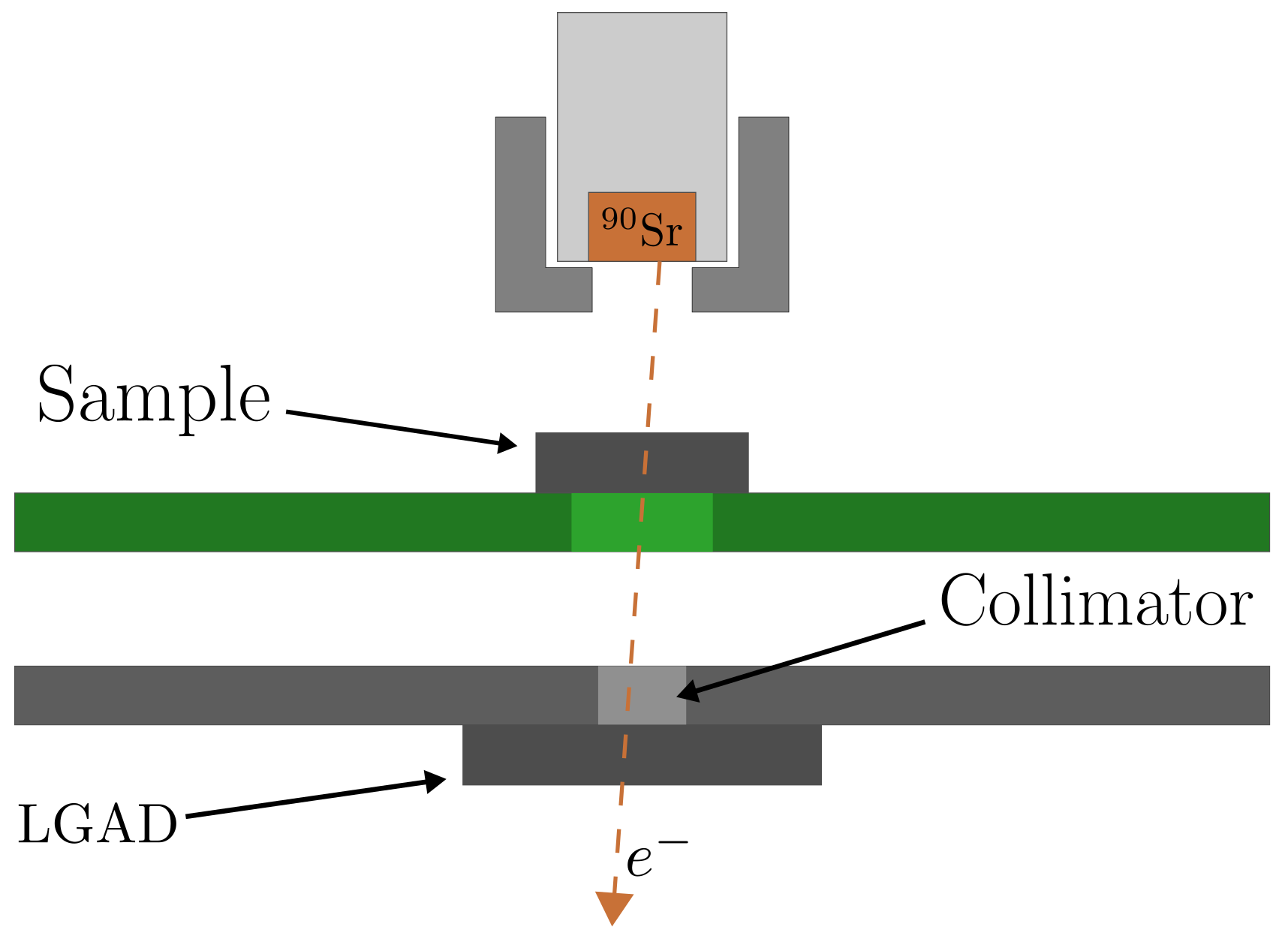}
	\caption[Schematic depiction of the $\Sr$ measurement setup]{
		Schematic depiction of the $\Sr$ measurement setup. The relevant
		distances and sizes are not to scale.
	}
	\label{fig:Sr90_setup}
\end{figure}

Time resolution was also determined by using electrons from a $\Sr$ source. A
schematic of the setup used in this case is shown in
figure~\ref{fig:Sr90_setup}, which follows a similar layout that was used
in~\cite{Kramberger_2019_3Dsensor_timing}
and~\cite{Kramberger_2020_LGAD_annealing_effects}. The reference signal is
provided by a second silicon detector mounted behind the sample. For a minimal
impact on the overall time resolution of the system, a thin Low Gain Avalanche
Detector~(LGAD) with a pad size of $\SI{1}{\mm}\times\SI{1}{\mm}$ and a time
resolution of around \SI{30}{\ps}, mounted on a timing board developed by
University of California Santa Cruz~\cite{Cartiglia_2017_16ps_timing_system},
was used for this purpose. A collimator is positioned in front of the reference
LGAD to only select electrons that pass through the device under test and create
a sufficient amount of charge in the reference detector, filtering out electrons
from the lower end of the energy spectrum that do not behave as MIPs.

The acquisition was triggered on coincident signals in both channels. Due to the
small pixel size in the RD50-MPW2, the hit rate was limited to
\SI{1}{\min\tothe{-1}} using an \SI{18}{\mega\becquerel} source. To eliminate
any time walk effects on the reference detector, constant fraction
discrimination at \SI{20}{\percent} of the maximum was used in the offline LGAD
signal analysis. The time resolution is again obtained as the standard deviation
of the $\mathrm{ToA}$ distribution. In this case, an extra contribution to the
measured time resolution arises from the reference detector
$\sigma_\mathrm{meas}^2=\sigma_\mathrm{MPW2}^2+\sigma_\mathrm{LGAD}^2$. Due to
the markedly better time resolution of the LGAD, the second term can be
neglected.

\begin{figure}
	\centering
	\begin{subfigure}{0.4\textwidth}
		\includegraphics[width=\textwidth]{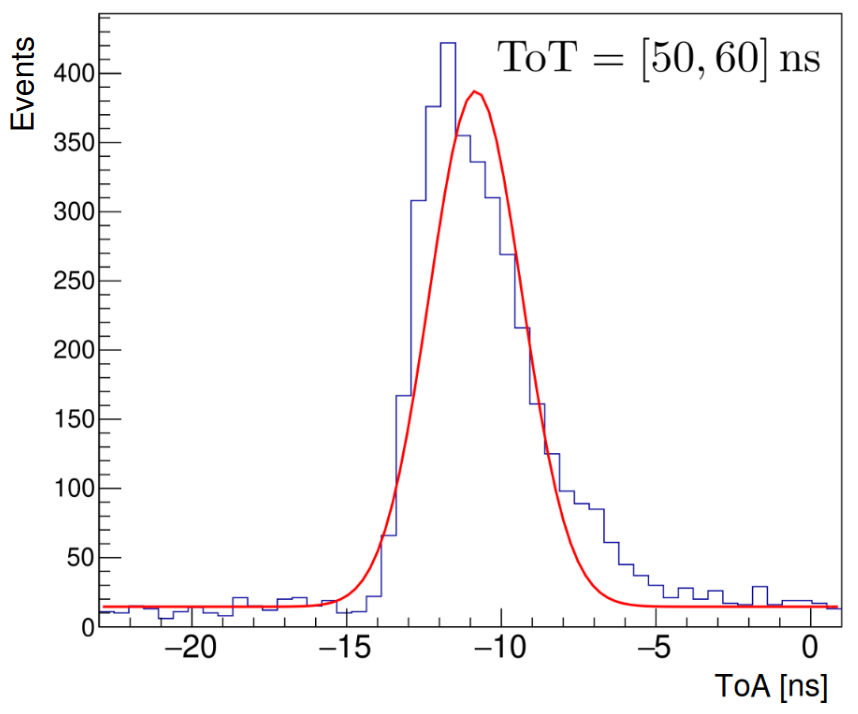}
		\caption{Unirradiated sample}
		\label{fig:Sr90_fit_irr0}
	\end{subfigure}
	\qquad
	\begin{subfigure}{0.4\textwidth}
		\includegraphics[width=\textwidth]{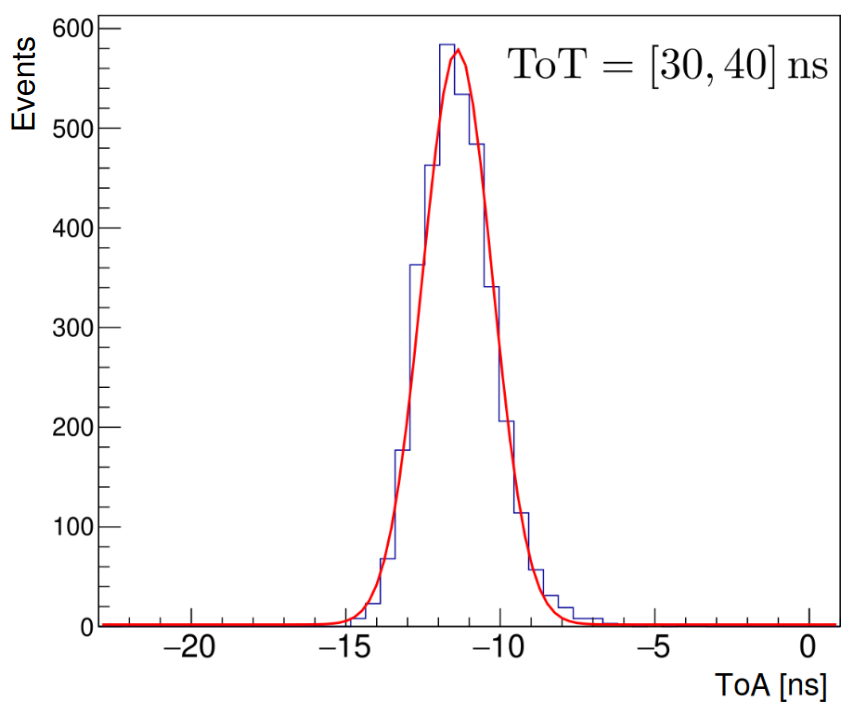}
		\caption{\fluence irradiated sample}
		\label{fig:Sr90_fit_irr5e14}
	\end{subfigure}
	\caption[$\mathrm{ToA}$ distributions for two bins in $\Sr$ measurements]{
		$\mathrm{ToA}$ distributions obtained with $\Sr$ measurements for two
		bins of comparable average arrival times. The edges of the
		$\mathrm{ToT}$ bins are noted in the plots. A Gaussian fit is performed
		to extract the time resolution.
	}
	\label{fig:Sr90_fit}
\end{figure}

\begin{figure}
	\centering
	\begin{subfigure}{0.495\textwidth}
		\includegraphics[width=\textwidth]{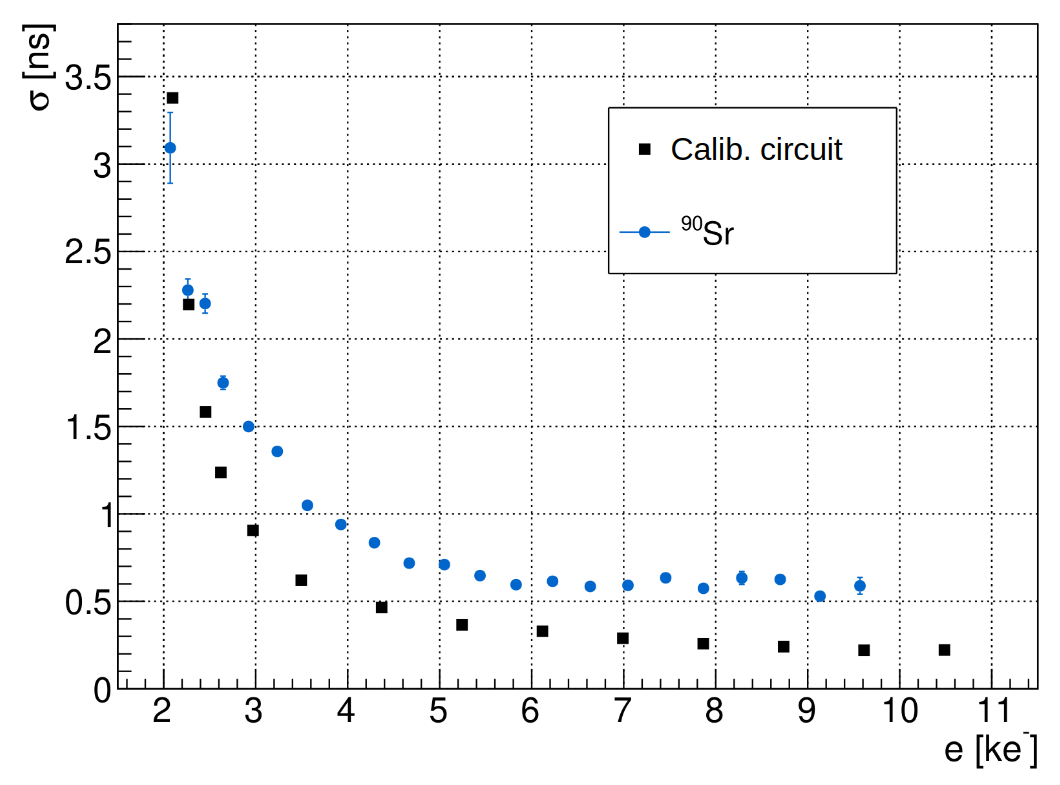}
		\caption{Unirradiated sample}
	\end{subfigure}
	\begin{subfigure}{0.495\textwidth}
		\includegraphics[width=\textwidth]{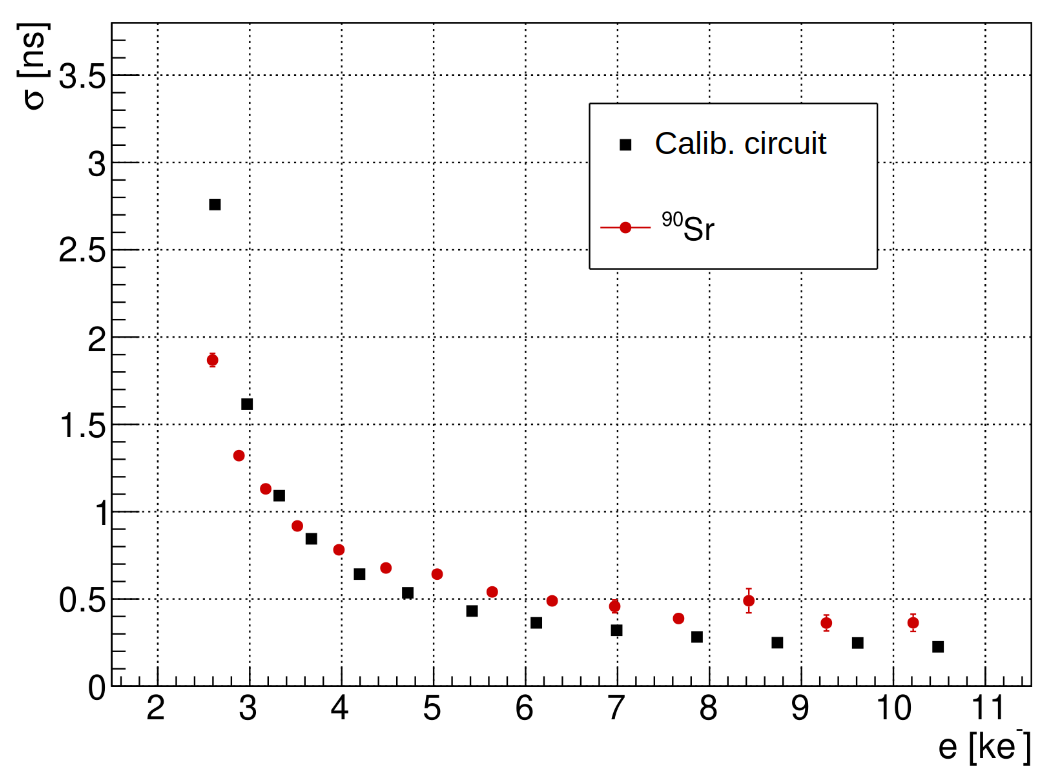}
		\caption{\fluence irradiated sample}
	\end{subfigure}
	\caption[Time resolution vs.\ collected charge for $\Sr$ measurements]{
		Time resolution vs.\ collected charge obtained with the $\Sr$ setup at a
		threshold of \SI{1000}{\mV}. Results using the calibration circuit are
		shown for comparison. Error bars indicate uncertainties of the Gaussian
		fit standard deviations.
	}
	\label{fig:Sr90_sigma}
\end{figure}

Since electrons from the $\Sr$ source deposit a variable amount of charge within
the pixel, the dependence of the time resolution on the collected charge is
determined by sorting all measured coincidence events by their $\mathrm{ToT}$
value of the comparator output and binning them into \SI{10}{\ns} wide bins.
Within each bin, the distribution of arrival times is fitted with a Gaussian
curve with its standard deviation representing the time resolution (see
figure~\ref{fig:Sr90_fit}). Results at a threshold of \SI{1000}{\mV} are
presented in figure~\ref{fig:Sr90_sigma}, where the collected charge for each
point has been obtained from the central $\mathrm{ToT}$ value of the respective
bin. The time resolution obtained with the $\Sr$ setup is worse than in the TCT
cases. For the unirradiated sample, the asymptotic time resolution at large
collected charge values is around \SI{600}{\ps}, whereas for the irradiated
sample, the time resolution improves to approx.\ \SI{360}{\ps} at charges of
\SI{10}{\kilo\electron}. These results suggest that irradiating the sample with
reactor neutrons to \fluence improves its timing performance. The origin of this
improvement can be seen in figure~\ref{fig:Sr90_fit}, where a large excess of
prolonged pixel responses is seen in the distribution for the unirradiated
sample bin. These large tails, which are most prominent at lower values of
$\mathrm{ToT}$, skew the bin distributions and widen the Gaussian fits,
resulting in larger values of the time resolution for the unirradiated sample.

Given that with irradiation similar performance improvements are not present in
the laser and calibration circuit measurements, the underlying cause of this
effect cannot be a consequence of the pixel electronics or signal processing,
but rather of the charge carrier creation and collection processes specific to
$\Sr$. Since the pixel is biased from the top of the chip (see pixel cross
section in~\cite{Hiti_2021_MPW2_timeWalk}), this results in curved electric
field lines and low field regions near the border of the depletion region, where
charge collection is presumed to be slow. This is confirmed by Backside-TCT
results in \cref{fig:attenuated_in_pixel_compare}, where charge created on the
edge of the pixel (corresponding to lower $\mathrm{ToT}$ values in the in-pixel
scan) produces signals with a larger delay than the same amount of charge
created in the pixel center, where collection is most efficient. The average
value of these delays can be up to a couple of \unit{\ns}, consistent with
observations in \cref{fig:Sr90_fit_irr0}. In addition, since the pixel is not
fully depleted at a bias voltage of \SI{100}{\V}, some amount of charge created
in the undepleted bulk reaches the depletion region via diffusion, subsequently
being collected on the electrode and contributing to the induced signal. Since
these delayed events cannot be filtered out in the $\Sr$ measurements, they
contribute to the time resolution calculation and worsen the results. After
irradiation, the charge carrier lifetime decreases due to deep energy levels
accelerating the charge carrier
recombination~\cite{Gaubas_2018_SiDefectTransformations}. As a result, less
charge from the slower component is able to reach the depleted
region~\cite{Kramberger_2019_HighCC} and produce a delayed pixel response, thus
essentially eliminating the delayed events in the bin distributions of the
irradiated sample (figure~\ref{fig:Sr90_fit_irr5e14}) and improving the time
resolution.

\section{Conclusion}
Timing properties were determined for active pixels of the RD50-MPW2 monolithic
prototype detector manufactured in \SI{150}{\nm} HV-CMOS technology.
Measurements at \SI{100}{\V} bias voltage with typical chip configuration and
threshold settings, and at deposited charge amounts close to those of MIPs, show
a minimum time resolution of approximately \SI{220}{\ps} measured with
Backside-TCT and \SI{300}{\ps} with Edge-TCT. Position sensitive scans in both
orientations indicate a uniform performance across the central pixel volume with
worse performance on the edges due to charge sharing and less efficient
collection. No significant performance differences were seen with the TCT method
when measuring a sample irradiated to \fluence. Measurements done with $\Sr$
give a worse time resolution of around \SI{600}{\ps} for the unirradiated
sample, which was attributed to events where charge is deposited on the
outermost parts of the pixel and charge collection is prolonged. Since charge
carriers recombine faster after irradiation, less delayed events are able to
produce a pixel response, thus improving the resolution.

Within the scope of characterizing the RD50-MPW2's timing performance, the TCT
method was utilized for the first time, both in Backside- and Edge-TCT
configurations, to determine the time resolution of monolithic detectors. In
comparison to $\Sr$ or test beam campaigns, two commonly used methods, time
resolution can be acquired with TCT in a shorter amount of time and,
additionally, also provides an insight into timing performance with positional
sensitivity on a sub-pixel level. Using TCT is therefore a suitable method for
obtaining time resolution information quickly and in a relatively simple
laboratory setting, giving it a promising future for applications in
characterizing time resolution of monolithic particle detectors.


\bibliographystyle{JHEP}
\bibliography{bibliography}

\providecommand{\href}[2]{#2}\begingroup\raggedright\begin{thebibliography}{10}

\bibitem{Sadrozinski_2017_4D_particle_tracking}
H.F.-W.~Sadrozinski, A.~Seiden and N.~Cartiglia, \emph{{4D} tracking with
  ultra-fast silicon detectors},
  \href{https://doi.org/10.1088/1361-6633/aa94d3}{\emph{Rep. Prog. Phys.}
  {\bfseries 81} (2018) 026101}.

\bibitem{Aberle_2020_TDR_HL-LHC}
I.~Béjar~Alonso, O.~Brüning, P.~Fessia, M.~Lamont, L.~Rossi, L.~Tavian
  et~al., eds., \emph{High-Luminosity Large Hadron Collider (HL-LHC): Technical
  design report}, CERN Yellow Reports: Monographs, CERN, Geneva (2020),
  \href{https://doi.org/10.23731/CYRM-2020-0010}{10.23731/CYRM-2020-0010}.

\bibitem{Abada_2019_FCC}
A.~Abada, M.~Abbrescia, S.S.~AbdusSalam, I.~Abdyukhanov,
  J.~Abelleira~Fernandez, A.~Abramov et~al., \emph{{FCC-hh}: The hadron
  collider}, \href{https://doi.org/10.1140/epjst/e2019-900087-0}{\emph{Eur.
  Phys. J. Spec. Top.} {\bfseries 228} (2019) 755}.

\bibitem{ATLAS_2012_PhaseII_upgrade_LOI}
{ATLAS Collaboration}, \emph{Letter of intent for the {Phase-II} upgrade of the
  {ATLAS} experiment},  Tech. Rep.
  \href{https://cds.cern.ch/record/1502664}{CERN-LHCC-2012-022, LHCC-I-023},
  CERN, Geneva (2012).

\bibitem{ATLAS_2020_HGTD_TDR}
{ATLAS Collaboration}, \emph{Technical design report: A {High-Granularity
  Timing Detector} for the {ATLAS} {Phase-II} upgrade},  Tech. Rep.
  \href{https://cds.cern.ch/record/2719855}{CERN-LHCC-2020-007, ATLAS-TDR-031},
  CERN, Geneva (2020).

\bibitem{Agapopoulou_2022_HGTD_LGAD_beam_tests}
C.~Agapopoulou, S.~Alderweireldt, S.~Ali, M.K.~Ayoub, D.~Benchekroun,
  L.~Castillo~García et~al., \emph{Performance in beam tests of irradiated low
  gain avalanche detectors for the {ATLAS} {High Granularity Timing Detector}},
  \href{https://doi.org/10.1088/1748-0221/17/09/P09026}{\emph{J. Instrum.}
  {\bfseries 17} (2022) P09026}.

\bibitem{Betancourt_2022_3D_detector_timeResolution}
C.~Betancourt, D.~De~Simone, G.~Kramberger, M.~Manna, G.~Pellegrini and
  N.~Serra, \emph{Time resolution of an irradiated {3D} silicon pixel
  detector},
  \href{https://doi.org/10.3390/instruments6010012}{\emph{Instruments}
  {\bfseries 6} (2022) 12}.

\bibitem{Iacobucci_2022_BiCMOSHexagonalPrototype_timeResolution}
G.~Iacobucci, L.~Paolozzi, P.~Valerio, T.~Moretti, F.~Cadoux, R.~Cardarelli
  et~al., \emph{Efficiency and time resolution of monolithic silicon pixel
  detectors in {SiGe} {BiCMOS} technology},
  \href{https://doi.org/10.1088/1748-0221/17/02/P02019}{\emph{J. Instrum.}
  {\bfseries 17} (2022) P02019}.

\bibitem{Peric_2007_DMAPS}
I.~Perić, \emph{A novel monolithic pixelated particle detector implemented in
  high-voltage {CMOS} technology},
  \href{https://doi.org/10.1016/j.nima.2007.07.115}{\emph{Nucl. Instrum.
  Methods Phys. Res. A} {\bfseries 582} (2007) 876}.

\bibitem{Kramberger_2010_EdgeTCT}
G.~Kramberger, V.~Cindro, I.~Mandić, M.~Mikuž, M.~Milovanović, M.~Zavrtanik
  et~al., \emph{Investigation of irradiated silicon detectors by {Edge-TCT}},
  \href{https://doi.org/10.1109/TNS.2010.2051957}{\emph{IEEE Trans. Nucl. Sci.}
  {\bfseries 57} (2010) 2294}.

\bibitem{Zhang_2020_RD50MPW2}
C.~Zhang, G.~Casse, N.~Massari, E.~Vilella and J.~Vossebeld, \emph{Development
  of {RD50-MPW2}: a high-speed monolithic {HV-CMOS} prototype chip within the
  {CERN-RD50} collaboration},  in \emph{PoS (TWEPP2019)}, vol.~370, p.~045,
  2020, \href{https://doi.org/10.22323/1.370.0045}{DOI}.

\bibitem{Vilella_2022_RD50MPW_developments}
E.~Vilella, \emph{Development of high voltage-{CMOS} sensors within the
  {CERN-RD50} collaboration},
  \href{https://doi.org/10.1016/j.nima.2022.166826}{\emph{Nucl. Instrum.
  Methods Phys. Res. A} {\bfseries 1034} (2022) 166826}.

\bibitem{Snoj_2012_TRIGA}
L.~Snoj, G.~Žerovnik and A.~Trkov, \emph{Computational analysis of irradiation
  facilities at the {JSI} {TRIGA} reactor},
  \href{https://doi.org/10.1016/j.apradiso.2011.11.042}{\emph{Appl. Radiat.
  Isot.} {\bfseries 70} (2012) 483}.

\bibitem{Ambrozic_2017_TRIGA_dose}
K.~Ambrožič, G.~Žerovnik and L.~Snoj, \emph{Computational analysis of the
  dose rates at {JSI} {TRIGA} reactor irradiation facilities},
  \href{https://doi.org/10.1016/j.apradiso.2017.09.022}{\emph{Appl. Radiat.
  Isot.} {\bfseries 130} (2017) 140}.

\bibitem{MarcoHernandez_2021_DMAPS_developments_at_RD50}
R.~Marco~Hernández, \emph{Latest depleted {CMOS} sensor developments in the
  {CERN} {RD50} collaboration},  in \emph{JPS Conf. Proc.}, vol.~34, p.~010008,
  2021, \href{https://doi.org/10.7566/JPSCP.34.010008}{DOI}.

\bibitem{Liu_2017_Caribou}
H.~Liu, M.~Benoit, H.~Chen, K.~Chen, F.A.~Di~Bello, G.~Iacobucci et~al.,
  \emph{Development of a modular test system for the silicon sensor {R\&D} of
  the {ATLAS} upgrade},
  \href{https://doi.org/10.1088/1748-0221/12/01/P01008}{\emph{J. Instrum.}
  {\bfseries 12} (2017) P01008}.

\bibitem{Vanat_2020_Caribou}
T.~Vanat, \emph{Caribou -- a versatile data acquisition system},
  \href{https://doi.org/10.22323/1.370.0100}{\emph{PoS} {\bfseries TWEPP2019}
  (2020) 100}.

\bibitem{Hiti_2021_MPW2_timeWalk}
B.~Hiti, V.~Cindro, A.~Gorišek, M.~Franks, R.~Marco-Hern\'andez, G.~Kramberger
  et~al., \emph{Characterisation of analogue front end and time walk in {CMOS}
  active pixel sensor},
  \href{https://doi.org/10.1088/1748-0221/16/12/P12020}{\emph{J. Instrum.}
  {\bfseries 16} (2021) P12020}.

\bibitem{Mandic_2022_MPW_passive_pixel_results}
I.~Mandić, V.~Cindro, J.~Debevc, A.~Gorišek, B.~Hiti, G.~Kramberger et~al.,
  \emph{Study of neutron irradiation effects in depleted {CMOS} detector
  structures}, \href{https://doi.org/10.1088/1748-0221/17/03/P03030}{\emph{J.
  Instrum.} {\bfseries 17} (2022) P03030}.

\bibitem{Particulars_homePage}
``{Particulars, Advanced measurement systems, Ltd}.''
\newblock \href{https://www.particulars.si/}{https://www.particulars.si/}.

\bibitem{Cartiglia_2014_UFSDperformance}
N.~Cartiglia, M.~Baselga, G.~Dellacasa, S.~Ely, V.~Fadeyev, Z.~Galloway et~al.,
  \emph{Performance of ultra-fast silicon detectors},
  \href{https://doi.org/10.1088/1748-0221/9/02/C02001}{\emph{J. Instrum.}
  {\bfseries 9} (2014) C02001}.

\bibitem{tsolantaMPW2}
C.~Tsolanta, \emph{Characterisation and time resolution measurements of the
  {RD50-MPW2} monolithic silicon pixel sensor},  Master's thesis, University of
  Amsterdam, 2022,
  \href{https://scripties.uba.uva.nl/search?id=record\_50779}{https://scripties.uba.uva.nl/search?id=record\_50779}.

\bibitem{Kramberger_2019_3Dsensor_timing}
G.~Kramberger, V.~Cindro, D.~Flores, S.~Hidalgo, B.~Hiti, M.~Manna et~al.,
  \emph{Timing performance of small cell {3D} silicon detectors},
  \href{https://doi.org/10.1016/j.nima.2019.04.088}{\emph{Nucl. Instrum.
  Methods Phys. Res. A} {\bfseries 934} (2019) 26}.

\bibitem{Kramberger_2020_LGAD_annealing_effects}
G.~Kramberger, V.~Cindro, A.~Howard, \v{Z}. Kljun, I.~Mandić and M.~Mikuž,
  \emph{Annealing effects on operation of thin {Low Gain Avalanche Detectors}},
  \href{https://doi.org/10.1088/1748-0221/15/08/P08017}{\emph{J. Instrum.}
  {\bfseries 15} (2020) P08017}.

\bibitem{Cartiglia_2017_16ps_timing_system}
N.~Cartiglia, A.~Staiano, V.~Sola, R.~Arcidiacono, R.~Cirio, F.~Cenna et~al.,
  \emph{Beam test results of a 16 ps timing system based on ultra-fast silicon
  detectors}, \href{https://doi.org/10.1016/j.nima.2017.01.021}{\emph{Nucl.
  Instrum. Methods Phys. Res. A} {\bfseries 850} (2017) 83}.

\bibitem{Gaubas_2018_SiDefectTransformations}
E.~Gaubas, T.~Ceponis, L.~Deveikis, D.~Meskauskaite, J.~Pavlov, V.~Rumbauskas
  et~al., \emph{Anneal induced transformations of defects in hadron irradiated
  {Si} wafers and {Schottky} diodes},
  \href{https://doi.org/10.1016/j.mssp.2017.11.035}{\emph{Mater. Sci. Semicond.
  Process.} {\bfseries 75} (2018) 157}.

\bibitem{Kramberger_2019_HighCC}
G.~Kramberger, \emph{Reasons for high charge collection efficiency of silicon
  detectors at {HL-LHC} fluences},
  \href{https://doi.org/10.1016/j.nima.2018.08.034}{\emph{Nucl. Instrum.
  Methods Phys. Res. A} {\bfseries 924} (2019) 192}.

\end{thebibliography}\endgroup

% Please avoid comments such as "For a review'', "For some examples",
% "and references therein" or move them in the text. In general,
% please leave only references in the bibliography and move all
% accessory text in footnotes.

\end{document}